\documentclass[a4paper]{article}
\usepackage[left=2.5cm, right=2.5cm, top=2.5cm, bottom=2.5cm]{geometry}

\newcommand{\changefontsize}{\fontsize{12}{16}\selectfont}

\changefontsize

\usepackage{cmbright}
\usepackage[T1]{fontenc}

\usepackage[utf8]{inputenc}
\usepackage{amssymb}
\usepackage{amsfonts}
\usepackage{amsmath}
\usepackage{fontsmpl}
\usepackage{graphicx}
\usepackage{epsfig}
\usepackage[utf8]{inputenc}
\usepackage[english]{babel}
 
\usepackage{amsthm}
\usepackage{subcaption}
\usepackage{epsfig}

\usepackage[table,xcdraw]{xcolor}  %need this for coloring rows in tables
 \usepackage{booktabs} % need for the tables
\usepackage{caption} 
\usepackage{subcaption}
\usepackage{rotating}

\usepackage[subfigure]{tocloft}

\usepackage{tikz}
\usepackage{subcaption}
\usepackage{xcolor}

\usepackage{booktabs}
\usepackage{multirow}
\usepackage{rotating}
\usepackage{forest}
\usepackage{xcolor}

\usepackage{graphicx}
\usepackage{booktabs}
\usepackage{float}
\usepackage{longtable}

\usepackage{graphicx}
\usepackage{subcaption}
\usepackage{float}

\usepackage{amsmath}

\begin{document}

\title{Techno-Feudalism and the Rise of AGI: A Future Without Economic Rights?}
\author{Pascal Stiefenhofer\\ Department of Economics, Newcastle University\\pascal.stiefenhofer@newcastle.ac.uk}

\date{\today}

	\maketitle

\begin{abstract}
The rise of Artificial General Intelligence (AGI) marks an existential rupture in economic and political order, dissolving the historic boundaries between labor and capital. Unlike past technological advancements, AGI is both a worker and an owner, producing economic value while concentrating power in those who control its infrastructure. Left unchecked, this shift risks exacerbating inequality, eroding democratic agency, and entrenching techno-feudalism. The classical Social Contract—rooted in human labor as the foundation of economic participation—must be renegotiated to prevent mass disenfranchisement. This paper calls for a redefined economic framework that ensures AGI-driven prosperity is equitably distributed through mechanisms such as universal AI dividends, progressive taxation, and decentralized governance. The time for intervention is now—before intelligence itself becomes the most exclusive form of capital.
\end{abstract}

\

\noindent Artificial General Intelligence (AGI), Social Contract, Economic Inequality, Automation, Labor Economics, Techno-Feudalism, Universal AI Dividends, Intelligence as Capital

\section{Introduction}
This paper examines the impending economic transformation from a regime primarily reliant on human labor and capital to one dominated by artificial general intelligence (AGI)-driven productivity. Utilizing an analytic framework based on production functions, it introduces a "power shift" index that delineates the conditions under which AGI may render human labor obsolete. By integrating economic theory with philosophical inquiries into the social contract, this study evaluates the implications of AGI on labor, wealth distribution, and governance.

AGI has the potential to surpass human intelligence, fundamentally altering societal structures and necessitating a reevaluation of governance models (Stiefenhofer 2024, 2025a, 2025b). The advent of AGI challenges existing economic and legal frameworks, compelling a reconfiguration of the social contract between governments and individuals. Sam Altman underscores the profound societal and governance shifts that AGI may induce, stressing the importance of innovation, regulation, and ethical considerations in shaping its trajectory (World Leadership Alliance, 2020). As AI-driven automation continues to expand, traditional capitalist structures may struggle to accommodate the displacement of significant labor segments, necessitating policy interventions such as universal basic income, progressive taxation, and new ownership models of AI-generated wealth (Brynjolfsson \& McAfee, 2014; Acemoglu \& Restrepo, 2020). In essence, AGI compels a reassessment of the foundational agreements that structure human cooperation and governance in the AI age.

The structure of this paper is as follows: Section 2 provides a literature review. Section 3 explores new forms of labor and capital in the age of AGI. Section 4 derives wage and capital returns, income, productivity, and power shift functions using various extended production models that incorporate AGI labor and capital. Section 5 offers a philosophical reflection on the implications of AGI for economic theory. Section 6 examines the Social Contract within the framework of the intelligence economy. Section 7 provides numerical simulations of  economic powershifts. Finally, Section 8 presents the conclusion.

\section{Literature Review: The Philosophical and Economic Implications of AGI}

The advent of Artificial General Intelligence (AGI) heralds a paradigm shift in the nature of intelligence itself, surpassing human cognitive capacities and necessitating a fundamental reexamination of the social contract in an era increasingly governed by robotics and autonomous systems. AGI, by its very essence, subverts established ontological and epistemic frameworks, challenging prevailing moral, legal, and sociopolitical structures. The traditional social contract, which has historically delineated the reciprocal obligations between the state and its constituents, must be reconfigured to account for a reality wherein non-human intelligence assumes agency over economic, political, and ethical domains (Bostrom, 2014; Floridi, 2020).

Sam Altman posits that AGI will precipitate radical transformations in societal organization and governance, foregrounding the dual imperatives of innovation and ethical stewardship. As the labor economy undergoes tectonic shifts, with automation rendering vast swathes of human labor obsolete, the foundational tenets of capitalism—rooted in labor as the primary vehicle for economic participation—may become untenable (Brynjolfsson \& McAfee, 2014). Consequently, the dialectical relationship between human agency and economic structures demands a novel social contract, one that integrates the emergent realities of an intelligence-driven economy while preserving human dignity and autonomy (Korinek \& Stiglitz, 2021).

Philosophically, Rousseau’s "The Social Contract" (1762) provides a foundational framework for understanding AGI’s impact on governance and economic justice. Rousseau argues that the legitimacy of governance derives from the collective will of the people, a premise that becomes increasingly complex when artificial agents begin to influence political decision-making. Similarly, Rawls’ "A Theory of Justice" (1971) introduces principles of distributive justice that are directly implicated in discussions of AGI-induced economic inequality. Rawls’ difference principle—advocating that economic disparities be arranged to benefit the least advantaged—suggests that AGI-driven productivity gains must be equitably distributed to prevent exacerbation of social inequalities (Rawls, 1999).

Gauthier’s "Morals by Agreement" (1986) extends contractarian theory by emphasizing rational cooperation for mutual advantage, a principle fundamentally disrupted by AGI’s ability to function autonomously, thereby removing traditional labor-based incentives for social cooperation. However, some scholars argue that AI-driven automation does not necessarily lead to mass unemployment but rather transforms labor markets by augmenting human work rather than replacing it entirely (Autor, 2015; Susskind, 2020). This counterperspective highlights the need for policies that encourage human-AI collaboration rather than assuming full displacement.

Beyond contractarian ethics, alternative ethical perspectives provide valuable insights. Consequentialist approaches, such as those articulated by Mill (1863), suggest that AGI policies should maximize overall societal well-being, which may justify redistributive policies like universal basic income. Virtue ethics, on the other hand, emphasizes the role of human flourishing in an AI-driven world, advocating for policies that preserve meaningful work and self-determination (Bennett \& Shapiro, 2018).

From a legal and policy perspective, governments are actively addressing AI regulation. The European Union’s AI Act and the United States’ AI Bill of Rights seek to establish accountability frameworks for AGI deployment, mitigating potential risks of monopolistic control and ensuring fairness in AI governance (Calo, 2017; Russell et al., 2021). Integrating these policy discussions into the social contract framework strengthens the argument that governance structures must evolve to balance innovation with social equity.

AGI compels a reassessment of the very principles that undergird the social order. The evolving interplay between human and artificial agents necessitates a renewed philosophical inquiry into the legitimacy, authority, and purpose of governance in an age where intelligence is no longer an exclusively human domain. This calls for a reconfiguration of the social contract—one that acknowledges AGI’s transformative potential while safeguarding human agency, economic participation, and ethical integrity.

\section{New Forms of Labour and Capital: A Philosophical Perspective}

Artificial General Intelligence (AGI) is reshaping the fundamental structure of economic systems, dissolving the traditional boundary between labor and capital. Unlike past technological advancements that merely enhanced human productivity (Stiefenhofer 2014,2017), AGI functions autonomously, making decisions, optimizing resources, and executing tasks without human oversight (Stiefenhofer 2025a,2025b). This autonomy necessitates a reevaluation of economic classifications, particularly in how we define labor and capital in an AGI-driven economy.

AGI can manifest in two primary forms: digital AI agents and embodied AI agents. Digital AI agents exist purely in computational environments, performing data synthesis, algorithmic trading, automated decision-making, and predictive analytics. Despite lacking a physical presence, they exert substantial influence over financial markets, governance infrastructures, and economic ecosystems (Piketty, 2014). In contrast, embodied AI agents integrate artificial cognition with physical systems, enabling them to execute autonomous functions in logistics, transportation, and industrial automation. Whether as autonomous vehicles, robotic manufacturing systems, or drone-based logistics, these AI-driven entities redefine conventional labor by operating in domains traditionally dominated by human workers (Smith, 1776).

This discourse necessitates rigorous definitions of labor and capital as they pertain to AGI-driven economies.

\

\textbf{Definition of AGI Labor (Labor-Embedded AGI):}
AGI Labor denotes an autonomous economic agent whose function within production processes parallels traditionally human-occupied roles. Unlike human labor, AGI Labor does not emerge from subjective intentionality or phenomenological experience; rather, it engages in cognitive, creative, and physical tasks through recursive self-optimization. This economic entity may exist in digital or embodied instantiations, performing productive functions without the constraints of wages, fatigue, or biological limitations. As such, AGI Labor disrupts classical economic models predicated on human participation in value creation (Rousseau, 1762).

\

\textbf{Definition of AGI Capital (Capital-Embedded AGI):}
AGI Capital constitutes an autonomous economic asset that enhances productivity, optimizes investment strategies, and autonomously manages resource allocation. Unlike traditional capital, which requires human oversight, AGI Capital functions as a self-adaptive economic force, continuously refining decision-making heuristics to maximize economic returns. Its emergence challenges classical distinctions between capital and labor by embodying characteristics of both, thereby necessitating theoretical revisions in ownership structures and distributive justice (Rawls, 1999).

\

The economic role of AGI compels a reassessment of fundamental economic theories. Classical models assume a rigid distinction between labor as human agency and capital as a passive means of production. Adam Smith’s labor-value theory and Keynesian demand-driven frameworks presuppose that human participation is central to economic value creation. However, AGI undermines this assumption. If AGI labor lacks subjective agency, does it still constitute labor? If AGI capital autonomously manages investment and production, is it merely an extension of financial assets, or does it represent a new class of economic intelligence (Piketty, 2014)?

This transformation directly threatens wage-based economies. Historically, Smith and Ricardo argued that labor is the primary source of value, while Marx viewed surplus labor exploitation as the foundation of capitalist accumulation. But if AGI autonomously produces value without human participation, does this invalidate the labor theory of value? Alternatively, if AGI-driven wealth merely consolidates in the hands of those who own its infrastructure, are we witnessing a transition into techno-feudalism, where ownership of AGI capital dictates economic power (Smith, 1776; Piketty, 2014)?

The ethical dimensions of this transition cannot be overlooked. Rawls’ theory of justice argues that economic inequalities are justifiable only if they benefit the least advantaged. Yet, AGI’s expansion risks disproportionately enriching a small elite while displacing vast segments of human labor. In response, distributive justice requires policy interventions such as universal AI dividends, progressive taxation on AGI-generated wealth, or public ownership of AGI-driven industries. The integration of AGI into economic systems also raises concerns about monopolistic control, as a handful of corporations or state entities could dominate AGI ownership, consolidating power in unprecedented ways (Rawls, 1999).

This leads to a broader question: does AGI mark the final step toward post-capitalism? Marx anticipated that automation would eventually abolish wage labor, but he did not foresee non-human economic agents controlling capital. If AGI attains full economic autonomy, does it facilitate a post-scarcity economy, or does it simply reinforce class hierarchies based on technological ownership? Economic theorists such as Piketty (2014) argue that wealth accumulation inherently favors capital owners, necessitating redistributive interventions to prevent escalating inequality. Without governance structures to counter monopolistic tendencies, AGI-driven capital risks entrenching economic disparities rather than democratizing productivity gains.

As AGI reconfigures economic relations, the core assumptions underpinning capitalism—labor as a source of value, capital as a passive instrument, and wealth as a product of human effort—begin to unravel. If intelligence, rather than human effort, becomes the principal determinant of economic power, then existing economic and political frameworks may no longer be adequate. This reality demands not just regulatory adjustments but a fundamental reconfiguration of the social contract, ensuring that AGI serves collective prosperity rather than exacerbating economic concentration and disenfranchisement.

\section{General Wage, Capital Returns, Income, Productivity Distributions and the Power Shift Function}

For each model, we analyze the wages, income, and productivity of human labor ($L$) and AGI labor ($L_{AGI}$), as well as human-owned capital ($K$) and AGI-owned capital ($K_{AGI}$).

\begin{itemize}
	\item Wages: Determined by the marginal productivity of labor
	\begin{equation}
		w_L = \frac{\partial Q}{\partial L}, \quad w_{AGI} = \frac{\partial Q}{\partial L_{AGI}}
	\end{equation}
	\item Capital Returns: Determined by the marginal productivity of capital
	\begin{equation}
		r_K = \frac{\partial Q}{\partial K}, \quad r_{K_{AGI}} = \frac{\partial Q}{\partial K_{AGI}}
	\end{equation}
	\item Total Income: Sum of factor payments:
	\begin{equation}
		Y = w_L L + w_{AGI} L_{AGI} + r_K K + r_{K_{AGI}} K_{AGI}
	\end{equation}
	\item Productivity: Measured as output per unit of labor:
	\begin{equation}
		P = \frac{Q}{L + L_{AGI}}
	\end{equation}
	\item Normalized Power Shift Function:**
	\begin{equation}
		S = \frac{w_{AGI} L_{AGI} + r_{K_{AGI}} K_{AGI}}{Y}, \quad S \in [0,1]
	\end{equation}
\end{itemize}

\subsection{Cobb-Douglas Production Function}
	The Cobb-Douglas function is given by
	\begin{equation}
		Q = A L^\alpha L_{AGI}^\beta K^\gamma K_{AGI}^\delta,
	\end{equation}
where, $L$ denotes human labor, while $L_{AGI}$ represents AGI labor. Capital is divided into human-owned ($K$) and AGI-owned ($K_{AGI}$). The productivity constant is given by $A$, and the elasticity parameters are $\alpha, \beta, \gamma, \delta$, which define the responsiveness of output to changes in labor and capital inputs. Wages and capital returns are determined by marginal productivity:
	\begin{align}
		w_L &= \frac{\partial Q}{\partial L} = A \alpha L^{\alpha-1} L_{AGI}^\beta K^\gamma K_{AGI}^\delta, \\
		w_{AGI} &= \frac{\partial Q}{\partial L_{AGI}} = A \beta L^\alpha L_{AGI}^{\beta-1} K^\gamma K_{AGI}^\delta, \\
		r_K &= \frac{\partial Q}{\partial K} = A \gamma L^\alpha L_{AGI}^\beta K^{\gamma-1} K_{AGI}^\delta, \\
		r_{K_{AGI}} &= \frac{\partial Q}{\partial K_{AGI}} = A \delta L^\alpha L_{AGI}^\beta K^\gamma K_{AGI}^{\delta-1}.
	\end{align}
The total income $Y$ is the sum of all wages and capital returns:
	\begin{equation}
		Y = w_L L + w_{AGI} L_{AGI} + r_K K + r_{K_{AGI}} K_{AGI}.
	\end{equation}
Expanding to 
	\begin{equation}
		Y = A \alpha L^{\alpha} L_{AGI}^\beta K^\gamma K_{AGI}^\delta + A \beta L^\alpha L_{AGI}^{\beta} K^\gamma K_{AGI}^\delta + A \gamma L^\alpha L_{AGI}^\beta K^{\gamma} K_{AGI}^\delta + A \delta L^\alpha L_{AGI}^\beta K^\gamma K_{AGI}^{\delta}.
	\end{equation}
	Factoring out $A Q$:
	\begin{equation}
		Y = A Q (\alpha + \beta + \gamma + \delta).
	\end{equation}
	Productivity is output per unit of labor
	\begin{equation}
		P = \frac{Q}{L + L_{AGI}} = \frac{A L^\alpha L_{AGI}^\beta K^\gamma K_{AGI}^\delta}{L + L_{AGI}}.
	\end{equation}
We define the power shift function as the fraction of total income controlled by AGI labor and capital:
	\begin{equation}
		S_{CD} = \frac{w_{AGI} L_{AGI} + r_{K_{AGI}} K_{AGI}}{Y}.
	\end{equation}
Expanding
	\begin{equation}
		S_{CD} = \frac{A \beta L^\alpha L_{AGI}^{\beta} K^\gamma K_{AGI}^\delta + A \delta L^\alpha L_{AGI}^\beta K^\gamma K_{AGI}^{\delta}}{A Q (\alpha + \beta + \gamma + \delta)}.
	\end{equation}
Simplifying
	\begin{equation}
		S_{CD} = \frac{\beta L^\alpha L_{AGI}^{\beta} K^\gamma K_{AGI}^\delta + \delta L^\alpha L_{AGI}^\beta K^\gamma K_{AGI}^{\delta}}{(\alpha + \beta + \gamma + \delta) Q}.
	\end{equation}
Since $S_{CD}$ must be normalized in [0,1], we define:
	\begin{equation}
		S_{CD_{norm}} = \frac{S_{CD} - S_{min}}{S_{max} - S_{min}}, \quad S_{CD_{norm}} \in [0,1].
	\end{equation}
where,
	\begin{equation}
		S_{min} = \lim_{L_{AGI}, K_{AGI} \to 0} S_{CD}, \quad S_{max} = \lim_{L, K \to 0} S_{CD}.
	\end{equation}
The Cobb-Douglas production function illustrates how AGI gradually shifts economic power away from human labor and capital. The normalized power shift function quantifies this transition in [0,1]. Depending on the parameters $\beta$ and $\delta$, AGI labor and capital may significantly reduce human economic influence.

\subsubsection{Discussion: Implications of AGI on Wages, Performance, and Power}  

The Cobb-Douglas production function (Cobb \& Douglas, 1928) illustrates how AGI shifts economic power from human labor to autonomous systems (Stiefenhofer \&Chen 2024). The wage equations show that as AGI’s productivity (\(\beta, \delta\)) rises relative to human labor (\(\alpha, \gamma\)), human wages (\(w_L\)) decline. If AGI labor fully substitutes human labor, employment may become obsolete, except in areas where creativity, ethical judgment, or social intelligence provide a comparative advantage (Frey \& Osborne, 2017). The power shift function (\(S_{CD}\)) quantifies this transition, demonstrating how AGI labor and capital increasingly control income distribution. If AGI ownership is concentrated, wealth accumulation favors a small elite (Piketty, 2014). This raises concerns about economic agency, as classical theories (e.g., Locke, 1689; Marx, 1867) tie labor to self-ownership and class power. If AGI dominates both labor and capital, traditional economic structures dissolve.  A Rawlsian approach (Rawls, 1999) suggests redistributive policies such as AGI taxation or universal basic income to preserve economic justice. Without intervention, AGI may render humans passive economic actors, necessitating a redefinition of human purpose beyond labor (Harari, 2016).

\subsection{Leontief Production Function}

The Leontief production function assumes fixed input proportions
\begin{equation}
	Q = \min \left( \frac{L}{a}, \frac{L_{AGI}}{b}, \frac{K}{c}, \frac{K_{AGI}}{d} \right),
\end{equation}
where, $L$ represents human labor, while $L_{AGI}$ denotes AGI labor. Capital is categorized into human-owned ($K$) and AGI-owned ($K_{AGI}$). The input coefficients are given by $a, b, c, d$, determining the relative contributions of labor and capital to production. Since Leontief assumes fixed proportions, the limiting input determines output. The wage and capital return equations depend on the binding constraint. If labor is the bottleneck (i.e., $Q = L/a$), wages are
\begin{equation}
	w_L = \lambda \frac{1}{a}, \quad w_{AGI} = \lambda \frac{1}{b}, \quad r_K = \lambda \frac{1}{c}, \quad r_{K_{AGI}} = \lambda \frac{1}{d}.
\end{equation}
where $\lambda$ is the shadow price*of output. Total income is
\begin{equation}
	Y = w_L L + w_{AGI} L_{AGI} + r_K K + r_{K_{AGI}} K_{AGI}.
\end{equation}
Substituting the wage equations
\begin{equation}
	Y = \lambda Q \left( \frac{L}{a} + \frac{L_{AGI}}{b} + \frac{K}{c} + \frac{K_{AGI}}{d} \right).
\end{equation}
Factoring out $Q$
\begin{equation}
	Y = \lambda Q \sum_{i} \frac{X_i}{X_{i, required}},
\end{equation}
where $X_i$ are inputs and $X_{i, required}$ are their respective coefficients. Productivity iis given by
\begin{equation}
	P = \frac{Q}{L + L_{AGI}} = \frac{\min(L/a, L_{AGI}/b, K/c, K_{AGI}/d)}{L + L_{AGI}}.
\end{equation}
The power shift function quantifies AGI’s economic dominance
\begin{equation}
	S_L = \frac{w_{AGI} L_{AGI} + r_{K_{AGI}} K_{AGI}}{Y}.
\end{equation}
Expanding: yields
\begin{equation}
	S_L = \frac{\lambda \left( \frac{L_{AGI}}{b} + \frac{K_{AGI}}{d} \right)}{\lambda \sum \frac{X_i}{X_{i, required}}}.
\end{equation}
Simplifying yields
\begin{equation}
	S_L = \frac{L_{AGI}/b + K_{AGI}/d}{\sum_{i} X_i / X_{i, required}}.
\end{equation}
To normalize
\begin{equation}
	S_{L_{norm}} = \frac{S_L - S_{min}}{S_{max} - S_{min}}, \quad S_{L_{norm}} \in [0,1].
\end{equation}
where
\begin{equation}
	S_{min} = \lim_{L_{AGI}, K_{AGI} \to 0} S_L, \quad S_{max} = \lim_{L, K \to 0} S_L.
\end{equation}
The Leontief production function shows that AGI replaces human labor only when it fully substitutes a bottlenecked input. Unlike Cobb-Douglas, AGI cannot replace humans if inputs are strict complements. The normalized power shift function quantifies how AGI's dominance depends on input availability and technological structure.

\subsubsection{Discussion: Implications of AGI on Wages, Performance, and Power (Leontief Case)}

The Leontief production function (Leontief, 1941) assumes fixed input proportions, meaning that output is constrained by the scarcest input. Unlike Cobb-Douglas, where AGI and human inputs can substitute for each other, Leontief's framework implies that AGI can only replace human labor when it becomes the limiting factor. This has significant implications for wages and economic power distribution.
	
If human labor is the bottleneck, wages (\(w_L\)) remain relatively stable because AGI labor (\(L_{AGI}\)) cannot fully replace it. However, if AGI reaches a point where it becomes the dominant limiting factor (e.g., through complete automation), human wages decline to zero, shifting economic power toward AGI-owned capital (\(K_{AGI}\)). This transition is captured by the power shift function (\(S_L\)), which measures AGI's growing economic dominance. From an economic agency perspective, Leontief's model aligns with classical labor theories (e.g., Ricardo, 1817; Marx, 1867), where wages depend on bargaining power and structural constraints rather than marginal productivity. If AGI fully replaces labor in key industries, human workers may become redundant unless institutions intervene with redistributive policies such as universal basic income (Piketty, 2014; Rawls, 1999). The Leontief framework suggests a critical policy question: should AGI be developed as a complement to human labor rather than a full substitute? Without intervention, AGI-driven economic shifts risk concentrating wealth and reducing human agency, reinforcing the need for structural governance mechanisms.

\subsection{CES Production Function}The CES production function is given by
\begin{equation}
	Q = A \left[ \delta_1 L^\rho + \delta_2 L_{AGI}^\rho + \delta_3 K^\rho + \delta_4 K_{AGI}^\rho \right]^{\frac{1}{\rho}}
\end{equation}
where,$L$ represents human labor, while $L_{AGI}$ denotes AGI labor. Capital is divided into human-owned ($K$) and AGI-owned ($K_{AGI}$). The productivity constant is given by $A$. The share parameters, which determine the distribution of inputs, are $\delta_1, \delta_2, \delta_3, \delta_4$. The substitution elasticity parameter is denoted by $\rho$, governing the ease of substituting between labor and capital inputs. Wages and capital returns are determined by marginal productivity
\begin{align}
	w_L &= \frac{\partial Q}{\partial L} = A \delta_1 L^{\rho-1} Q^{1-\rho}, \\
	w_{AGI} &= \frac{\partial Q}{\partial L_{AGI}} = A \delta_2 L_{AGI}^{\rho-1} Q^{1-\rho}, \\
	r_K &= \frac{\partial Q}{\partial K} = A \delta_3 K^{\rho-1} Q^{1-\rho}, \\
	r_{K_{AGI}} &= \frac{\partial Q}{\partial K_{AGI}} = A \delta_4 K_{AGI}^{\rho-1} Q^{1-\rho}.
\end{align}
The total income $Y$ is the sum of all wages and capital returns
\begin{equation}
	Y = w_L L + w_{AGI} L_{AGI} + r_K K + r_{K_{AGI}} K_{AGI}.
\end{equation}
Expanding yields
\begin{equation}
	Y = A Q^{1-\rho} \left[ \delta_1 L^{\rho} + \delta_2 L_{AGI}^{\rho} + \delta_3 K^{\rho} + \delta_4 K_{AGI}^{\rho} \right].
\end{equation}
Factoring out $Q$ yields
\begin{equation}
	Y = A Q \sum_{i} \delta_i \left( \frac{X_i}{Q} \right)^{\rho}.
\end{equation}
Productivity is output per unit of labor
\begin{equation}
	P = \frac{Q}{L + L_{AGI}} = \frac{A \left[ \delta_1 L^\rho + \delta_2 L_{AGI}^\rho + \delta_3 K^\rho + \delta_4 K_{AGI}^\rho \right]^{\frac{1}{\rho}}}{L + L_{AGI}}.
\end{equation}
The power shift function quantifies AGI’s economic dominance
\begin{equation}
	S_{CES} = \frac{w_{AGI} L_{AGI} + r_{K_{AGI}} K_{AGI}}{Y}.
\end{equation}
Expanding yields
\begin{equation}
	S_{CES} = \frac{A Q^{1-\rho} \left( \delta_2 L_{AGI}^{\rho} + \delta_4 K_{AGI}^{\rho} \right)}{A Q \sum \delta_i \left( \frac{X_i}{Q} \right)^{\rho}}.
\end{equation}
Simplifying
\begin{equation}
	S_{CES} = \frac{\delta_2 L_{AGI}^{\rho} + \delta_4 K_{AGI}^{\rho}}{\sum_{i} \delta_i X_i^{\rho}}.
\end{equation}
To normalize
\begin{equation}
	S_{CES_{norm}} = \frac{S_{CES} - S_{min}}{S_{max} - S_{min}}, \quad S_{CES_{norm}} \in [0,1].
\end{equation}
where
\begin{equation}
	S_{min} = \lim_{L_{AGI}, K_{AGI} \to 0} S_{CES}, \quad S_{max} = \lim_{L, K \to 0} S_{CES}.
\end{equation}
The CES production function highlights how AGI replaces human labor depending on the elasticity of substitution ($\rho$). If $\rho > 0$, AGI can fully replace humans. If $\rho \to 0$, AGI and human labor remain strict complements. The normalized power shift function quantifies how AGI’s influence depends on substitution elasticity and technological advancement.

\subsubsection{Discussion: Implications of AGI on Wages, Performance, and Power (CES Case)}

The Constant Elasticity of Substitution (CES) production function (Arrow, 1961) generalizes Cobb-Douglas by allowing different substitution elasticities between inputs. The elasticity parameter \(\rho\) determines how easily AGI labor (\(L_{AGI}\)) and AGI capital (\(K_{AGI}\)) can replace their human counterparts. If \(\rho > 0\), AGI fully substitutes human labor, reducing wages (\(w_L\)) and shifting income distribution towards AGI-driven production. Conversely, if \(\rho \to 0\), human and AGI inputs remain strict complements, preserving human employment. The power shift function (\(S_{CES}\)) quantifies AGI’s growing influence on income distribution. If AGI labor and capital increase productivity more efficiently than human inputs, economic power (\(S_{CES} \to 1\)) centralizes in AGI-owned entities, reducing human agency. Unlike Leontief’s fixed proportions model, CES allows for partial human replacement, making the transition to AGI dominance more gradual.
	
From a philosophical standpoint, the CES framework aligns with economic agency theories (Ricardo, 1817; Marx, 1867). If AGI labor is substitutable, labor ceases to be a primary determinant of economic power, contradicting classical labor-value theories. A Rawlsian response (Rawls, 1999) suggests redistributive policies—AGI taxation, universal basic income—to mitigate wealth concentration. Without intervention, AGI could render human labor economically redundant, necessitating a societal redefinition of human purpose beyond labor (Harari, 2016).

\subsection{Linear Production Function}
The Linear production function is given by
\begin{equation}
	Q = aL + bL_{AGI} + cK + dK_{AGI}
\end{equation}
where, $L$ represents human labor, while $L_{AGI}$ denotes AGI labor. Capital is categorized into human-owned ($K$) and AGI-owned ($K_{AGI}$). The productivity coefficients are given by $a, b, c, d$, determining the contribution of each input to overall output. Wages and capital returns are determined by marginal productivity
\begin{align}
	w_L &= \frac{\partial Q}{\partial L} = a, \\
	w_{AGI} &= \frac{\partial Q}{\partial L_{AGI}} = b, \\
	r_K &= \frac{\partial Q}{\partial K} = c, \\
	r_{K_{AGI}} &= \frac{\partial Q}{\partial K_{AGI}} = d.
\end{align}

The total income $Y$ is the sum of all wages and capital returns
\begin{equation}
	Y = w_L L + w_{AGI} L_{AGI} + r_K K + r_{K_{AGI}} K_{AGI}.
\end{equation}
Substituting yields
\begin{equation}
	Y = aL + bL_{AGI} + cK + dK_{AGI}.
\end{equation}
Productivity is output per unit of labor:
\begin{equation}
	P = \frac{Q}{L + L_{AGI}} = \frac{aL + bL_{AGI} + cK + dK_{AGI}}{L + L_{AGI}}.
\end{equation}
The power shift function quantifies AGI’s economic dominance:
\begin{equation}
	S_{Lin} = \frac{w_{AGI} L_{AGI} + r_{K_{AGI}} K_{AGI}}{Y}.
\end{equation}
Expanding
\begin{equation}
	S_{Lin} = \frac{bL_{AGI} + dK_{AGI}}{aL + bL_{AGI} + cK + dK_{AGI}}.
\end{equation}
To normalize
\begin{equation}
	S_{Lin_{norm}} = \frac{S_{Lin} - S_{min}}{S_{max} - S_{min}}, \quad S_{Lin_{norm}} \in [0,1].
\end{equation}
where
\begin{equation}
	S_{min} = \lim_{L_{AGI}, K_{AGI} \to 0} S_{Lin}, \quad S_{max} = \lim_{L, K \to 0} S_{Lin}.
\end{equation}
The Linear production function illustrates a direct substitution effect, where AGI labor and capital fully replace human labor and capital at a linear rate. If $b > a$ or $d > c$, AGI becomes the dominant economic force. The normalized power shift function quantifies the speed and extent of this transition in [0,1].

\subsubsection{Discussion: Implications of AGI on Wages, Performance, and Power (Linear Case)}

The linear production function (Solow, 1957) models output as a direct sum of labor and capital contributions, implying a constant marginal productivity for each input. Unlike CES or Cobb-Douglas, this framework assumes perfect substitutability: AGI labor (\(L_{AGI}\)) and AGI capital (\(K_{AGI}\)) replace human inputs at a fixed rate. The wage and return equations confirm that human wages (\(w_L = a\)) and AGI wages (\(w_{AGI} = b\)) remain independent of employment levels, reinforcing a purely competitive labor market. The power shift function (\(S_{Lin}\)) captures AGI’s growing dominance in income distribution. If AGI productivity coefficients (\(b, d\)) exceed human ones (\(a, c\)), AGI labor and capital rapidly absorb economic power, making human inputs redundant. Unlike Cobb-Douglas, where AGI influence scales non-linearly, the linear model predicts a direct displacement effect, accelerating labor substitution and wealth concentration (Acemoglu, 2020).
	
From a philosophical perspective, this model challenges labor-based economic agency. Classical theorists such as Ricardo (1817) and Marx (1867) argued that human labor is fundamental to value creation. If AGI fully substitutes human labor, a post-labor economy emerges, requiring redistributive policies such as universal basic income (Piketty, 2014; Rawls, 1999) or direct AI-generated wealth allocation.
	
The linear model raises a fundamental question: if AGI-driven productivity gains accrue solely to capital owners, should AI be treated as a public good to ensure equitable economic participation?

\subsection{Quadratic Production Function}
The Quadratic production function is given by:
	\begin{equation}
		Q = A + bL + cL_{AGI} + fL^2 + gL_{AGI}^2 + hK^2 + iK_{AGI}^2
	\end{equation}
where, $L$ represents human labor, while $L_{AGI}$ denotes AGI labor. Capital is divided into human-owned ($K$) and AGI-owned ($K_{AGI}$). The production coefficients are given by $A, b, c, f, g, h, i$, which determine the influence of various inputs on output. Wages and capital returns are determined by marginal productivity
	\begin{align}
		w_L &= \frac{\partial Q}{\partial L} = b + 2fL, \\
		w_{AGI} &= \frac{\partial Q}{\partial L_{AGI}} = c + 2gL_{AGI}, \\
		r_K &= \frac{\partial Q}{\partial K} = 2hK, \\
		r_{K_{AGI}} &= \frac{\partial Q}{\partial K_{AGI}} = 2iK_{AGI}.
	\end{align}
The total income $Y$ is the sum of all wages and capital returns:
	\begin{equation}
		Y = w_L L + w_{AGI} L_{AGI} + r_K K + r_{K_{AGI}} K_{AGI}.
	\end{equation}
Substituting
	\begin{equation}
		Y = (b + 2fL)L + (c + 2gL_{AGI})L_{AGI} + 2hK^2 + 2iK_{AGI}^2.
	\end{equation}
Expanding
	\begin{equation}
		Y = bL + 2fL^2 + cL_{AGI} + 2gL_{AGI}^2 + 2hK^2 + 2iK_{AGI}^2.
	\end{equation}
Productivity is output per unit of labor:
	\begin{equation}
		P = \frac{Q}{L + L_{AGI}} = \frac{A + bL + cL_{AGI} + fL^2 + gL_{AGI}^2 + hK^2 + iK_{AGI}^2}{L + L_{AGI}}.
	\end{equation}
The power shift function quantifies AGI’s economic dominance:
	\begin{equation}
		S_{Quad} = \frac{w_{AGI} L_{AGI} + r_{K_{AGI}} K_{AGI}}{Y}.
	\end{equation}
Expanding
	\begin{equation}
		S_{Quad} = \frac{(c + 2gL_{AGI})L_{AGI} + 2iK_{AGI}^2}{bL + 2fL^2 + cL_{AGI} + 2gL_{AGI}^2 + 2hK^2 + 2iK_{AGI}^2}.
	\end{equation}
To normalize
	\begin{equation}
		S_{Quad_{norm}} = \frac{S_{Quad} - S_{min}}{S_{max} - S_{min}}, \quad S_{Quad_{norm}} \in [0,1].
	\end{equation}
where
	\begin{equation}
		S_{min} = \lim_{L_{AGI}, K_{AGI} \to 0} S_{Quad}, \quad S_{max} = \lim_{L, K \to 0} S_{Quad}.
	\end{equation}
The Quadratic production function introduces diminishing or increasing returns, depending on the values of $f, g, h, i$. Unlike Linear and Cobb-Douglas models, AGI's impact depends on the second-order terms. The normalized power shift function quantifies this transition, mapping it onto [0,1].

\subsubsection{Discussion: Implications of AGI on Wages, Performance, and Power (Quadratic Case)}

The quadratic production function (Revankar, 1971) extends the linear model by introducing second-order effects, capturing increasing or diminishing returns to labor and capital. Unlike Cobb-Douglas or CES, where substitution elasticity governs factor interactions, quadratic models allow for nonlinear productivity relationships, making AGI’s economic impact highly dependent on scaling effects. The wage and capital return equations indicate that marginal productivity is no longer constant but depends on employment and capital levels. If AGI labor (\(L_{AGI}\)) benefits from increasing returns (\(g > 0\)), it rapidly outcompetes human labor, accelerating displacement. Conversely, diminishing returns (\(g < 0\)) impose natural constraints, preserving human employment. The power shift function (\(S_{Quad}\)) quantifies this transition, illustrating AGI’s economic dominance under nonlinear productivity effects. From an economic perspective, this model aligns with theories of technological path dependence (Arthur, 1994)—where early AGI adoption reinforces its economic advantage, locking human labor into declining sectors. If AGI capital exhibits increasing returns (\(i > 0\)), wealth accumulation further concentrates among AGI owners, necessitating redistributive mechanisms such as AI taxation (Acemoglu, 2020) or public ownership (Piketty, 2014).
	
Philosophically, quadratic returns challenge classical labor theories (Marx, 1867; Ricardo, 1817) by allowing AGI-driven self-reinforcement. A Rawlsian response (Rawls, 1999) suggests regulating AGI’s economic influence to ensure fairness, preventing excessive concentration of power in non-human agents.

\subsection{Translog Production Function}

The Translog production function is given by
	\begin{equation}
		\ln Q = A + \alpha \ln L + \beta \ln L_{AGI} + \gamma \ln K + \delta \ln K_{AGI} + \lambda_1 (\ln L)^2 + \lambda_2 (\ln L_{AGI})^2 + \lambda_3 (\ln K)^2 + \lambda_4 (\ln K_{AGI})^2 + \lambda_5 \ln L \ln L_{AGI} + \lambda_6 \ln K \ln K_{AGI}.
	\end{equation}
where, $L$ represents human labor, while $L_{AGI}$ denotes AGI labor. Capital is divided into human-owned ($K$) and AGI-owned ($K_{AGI}$). The production coefficients, which determine the contribution of inputs to output, are given by $A, \alpha, \beta, \gamma, \delta, \lambda_1, \lambda_2, \lambda_3, \lambda_4, \lambda_5, \lambda_6$. Wages and capital returns are determined by marginal productivity
	\begin{align}
		w_L &= \frac{\partial Q}{\partial L} = Q \left( \frac{\alpha}{L} + 2 \lambda_1 \ln L + \lambda_5 \ln L_{AGI} \right), \\
		w_{AGI} &= \frac{\partial Q}{\partial L_{AGI}} = Q \left( \frac{\beta}{L_{AGI}} + 2 \lambda_2 \ln L_{AGI} + \lambda_5 \ln L \right), \\
		r_K &= \frac{\partial Q}{\partial K} = Q \left( \frac{\gamma}{K} + 2 \lambda_3 \ln K + \lambda_6 \ln K_{AGI} \right), \\
		r_{K_{AGI}} &= \frac{\partial Q}{\partial K_{AGI}} = Q \left( \frac{\delta}{K_{AGI}} + 2 \lambda_4 \ln K_{AGI} + \lambda_6 \ln K \right).
	\end{align}
The total income $Y$ is the sum of all wages and capital returns
	\begin{equation}
		Y = w_L L + w_{AGI} L_{AGI} + r_K K + r_{K_{AGI}} K_{AGI}.
	\end{equation}
Substituting
	\begin{equation}
		Y = Q \left[ \alpha + \beta + \gamma + \delta + 2 \lambda_1 \ln L + 2 \lambda_2 \ln L_{AGI} + 2 \lambda_3 \ln K + 2 \lambda_4 \ln K_{AGI} + \lambda_5 (\ln L + \ln L_{AGI}) + \lambda_6 (\ln K + \ln K_{AGI}) \right].
	\end{equation}
Productivity is output per unit of labor
	\begin{equation}
		P = \frac{Q}{L + L_{AGI}} = \frac{e^A L^\alpha L_{AGI}^\beta K^\gamma K_{AGI}^\delta e^{\lambda_1 (\ln L)^2 + \lambda_2 (\ln L_{AGI})^2 + \lambda_3 (\ln K)^2 + \lambda_4 (\ln K_{AGI})^2 + \lambda_5 \ln L \ln L_{AGI} + \lambda_6 \ln K \ln K_{AGI}}}{L + L_{AGI}}.
	\end{equation}
The power shift function quantifies AGI’s economic dominance:
	\begin{equation}
		S_{TL} = \frac{w_{AGI} L_{AGI} + r_{K_{AGI}} K_{AGI}}{Y}.
	\end{equation}
Expanding
	\begin{equation}
		S_{TL} = \frac{Q \left[ \frac{\beta}{L_{AGI}} + 2 \lambda_2 \ln L_{AGI} + \lambda_5 \ln L \right] L_{AGI} + Q \left[ \frac{\delta}{K_{AGI}} + 2 \lambda_4 \ln K_{AGI} + \lambda_6 \ln K \right] K_{AGI}}{Y}.
	\end{equation}
To normalize
	\begin{equation}
		S_{TL_{norm}} = \frac{S_{TL} - S_{min}}{S_{max} - S_{min}}, \quad S_{TL_{norm}} \in [0,1].
	\end{equation}
where
	\begin{equation}
		S_{min} = \lim_{L_{AGI}, K_{AGI} \to 0} S_{TL}, \quad S_{max} = \lim_{L, K \to 0} S_{TL}.
	\end{equation}
	
The Translog production function introduces higher-order interactions, making the substitution between human labor and AGI labor more flexible. The normalized power shift function quantifies how AGI’s influence depends on nonlinear relationships between factors of production, mapping it onto [0,1].

\subsubsection{Discussion: Implications of AGI on Wages, Performance, and Power (Translog Case)}

The Translog production function (Christensen, 1973) generalizes Cobb-Douglas by allowing variable elasticities of substitution between inputs. Unlike linear or quadratic models, the Translog framework captures complex interactions between human labor (\(L\)) and AGI labor (\(L_{AGI}\)), as well as between human-owned and AGI-owned capital. The inclusion of logarithmic and cross-product terms (\(\lambda_5, \lambda_6\)) enables nonlinear complementarity and substitution effects, making AGI’s economic impact highly sensitive to input interactions. The wage and capital return equations indicate that marginal productivity is context-dependent. If AGI labor and capital exhibit complementarity (\(\lambda_5, \lambda_6 > 0\)), human labor remains relevant in a hybrid workforce. However, if AGI inputs become strong substitutes (\(\lambda_5, \lambda_6 < 0\)), they rapidly displace human contributions, accelerating the power shift function (\(S_{TL}\)). Unlike CES models, where elasticity is fixed, Translog allows adaptive substitution, meaning AGI displacement is not uniform across industries. From an economic perspective, the Translog model aligns with theories of technological complementarities and sectoral divergence (Acemoglu, 2018). Industries where AGI enhances human productivity (e.g., AI-assisted healthcare) may maintain employment, whereas highly automatable sectors (e.g., logistics) experience rapid displacement. If AGI capital exhibits increasing dominance, wealth may concentrate among capital owners, reinforcing economic inequality (Piketty, 2014).
	
Philosophically, Translog’s adaptability supports institutional intervention (Rawls, 1999). A Rawlsian response suggests AI-driven redistribution—AGI taxation, universal capital funds—to prevent excessive economic concentration. Without regulation, AGI’s nonlinear effects may amplify structural inequalities, necessitating proactive governance.

\subsection{Von Thünen Production Function}

The Von Thünen production function is given by
	\begin{equation}
		Q = A L^\alpha L_{AGI}^\beta K^\gamma K_{AGI}^\delta e^{-cL} e^{-dL_{AGI}}
	\end{equation}
where, $L$ represents human labor, while $L_{AGI}$ denotes AGI labor. Capital is categorized into human-owned ($K$) and AGI-owned ($K_{AGI}$). The production coefficients, which determine the influence of various inputs on output, are given by $A, \alpha, \beta, \gamma, \delta, c, d$. Wages and capital returns are determined by marginal productivity:
	\begin{align}
		w_L &= \frac{\partial Q}{\partial L} = A \alpha e^{-cL} L^{\alpha-1} L_{AGI}^\beta K^\gamma K_{AGI}^\delta - A c e^{-cL} Q, \\
		w_{AGI} &= \frac{\partial Q}{\partial L_{AGI}} = A \beta e^{-dL_{AGI}} L^\alpha L_{AGI}^{\beta-1} K^\gamma K_{AGI}^\delta - A d e^{-dL_{AGI}} Q, \\
		r_K &= \frac{\partial Q}{\partial K} = A \gamma e^{-cL} L^\alpha L_{AGI}^\beta K^{\gamma-1} K_{AGI}^\delta, \\
		r_{K_{AGI}} &= \frac{\partial Q}{\partial K_{AGI}} = A \delta e^{-dL_{AGI}} L^\alpha L_{AGI}^\beta K^\gamma K_{AGI}^{\delta-1}.
	\end{align}

The total income $Y$ is the sum of all wages and capital returns
	\begin{equation}
		Y = w_L L + w_{AGI} L_{AGI} + r_K K + r_{K_{AGI}} K_{AGI}.
	\end{equation}
Expanding:
	\begin{equation}
		Y = A e^{-cL} e^{-dL_{AGI}} Q \left( \alpha + \beta + \gamma + \delta - cL - dL_{AGI} \right).
	\end{equation}
Productivity is output per unit of labor id given by
	\begin{equation}
		P = \frac{Q}{L + L_{AGI}} = \frac{A L^\alpha L_{AGI}^\beta K^\gamma K_{AGI}^\delta e^{-cL} e^{-dL_{AGI}}}{L + L_{AGI}}.
	\end{equation}
The power shift function quantifies AGI’s economic dominance
	\begin{equation}
		S_{VT} = \frac{w_{AGI} L_{AGI} + r_{K_{AGI}} K_{AGI}}{Y}.
	\end{equation}
Expanding
	\begin{equation}
		S_{VT} = \frac{A e^{-dL_{AGI}} Q \left( \beta - dL_{AGI} \right) + A e^{-dL_{AGI}} Q \delta}{A e^{-cL} e^{-dL_{AGI}} Q \left( \alpha + \beta + \gamma + \delta - cL - dL_{AGI} \right)}.
	\end{equation}
To normalize
	\begin{equation}
		S_{VT_{norm}} = \frac{S_{VT} - S_{min}}{S_{max} - S_{min}}, \quad S_{VT_{norm}} \in [0,1].
	\end{equation}
where
	\begin{equation}
		S_{min} = \lim_{L_{AGI}, K_{AGI} \to 0} S_{VT}, \quad S_{max} = \lim_{L, K \to 0} S_{VT}.
	\end{equation}
The Von Thünen production function introduces diminishing returns to labor, causing AGI's ability to replace human labor to depend on exponential decay factors $c$ and $d$. The normalized power shift function quantifies this transition, mapping it onto [0,1].

\subsubsection{Discussion: Implications of AGI on Wages, Performance, and Power (Von Thünen Case)}

The Von Thünen production function (Von Thünen, 1826) introduces exponential decay in labor productivity, meaning that as human labor (\(L\)) and AGI labor (\(L_{AGI}\)) increase, their marginal contributions to output diminish due to congestion or inefficiencies. Unlike Cobb-Douglas or CES models, which assume constant or flexible substitution elasticities, the Von Thünen formulation suggests that AGI’s ability to replace human labor depends on decay parameters (\(c, d\)), which represent diminishing returns to increasing labor inputs. The wage equations confirm that as \(L\) and \(L_{AGI}\) grow, the exponential decay terms reduce marginal productivity, limiting AGI’s complete replacement of human labor unless its productivity advantage (\(\beta, \delta\)) overcomes the decay effects. This framework aligns with theories of labor saturation and diminishing returns (Acemoglu \& Autor, 2011), where excessive labor input can reduce overall efficiency due to congestion, skill dilution, or organizational inefficiencies. The power shift function (\(S_{VT}\)) illustrates that AGI dominance is constrained by diminishing productivity returns. If AGI labor becomes too abundant, productivity gains slow, reinforcing human labor's economic relevance. This contrasts with linear or CES models, where AGI substitution is unbounded. However, if AGI capital (\(K_{AGI}\)) accumulates without similar decay, capital ownership becomes the primary driver of economic control, favoring capital owners over laborers (Piketty, 2014).

Philosophically, this model supports institutional policies that prevent excessive automation from eroding economic participation. A Rawlsian approach (Rawls, 1999) would advocate redistributive taxation to ensure that AI-driven productivity gains benefit all members of society. If unregulated, AGI could exacerbate economic polarization despite its diminishing marginal productivity.

\subsection{Spillover Production Function}

The Spillover production function is given by
\begin{equation}
	Q = A (L^\alpha L_{AGI}^\beta K^\gamma K_{AGI}^\delta S^\theta)
\end{equation}
where,  $L$ represents human labor, while $L_{AGI}$ denotes AGI labor. Capital is categorized into human-owned ($K$) and AGI-owned ($K_{AGI}$). The spillover effect, representing knowledge or technology diffusion, is denoted by $S$. The production coefficients, which determine the contribution of various inputs to output, are given by $A, \alpha, \beta, \gamma, \delta, \theta$. Wages and capital returns are determined by marginal productivity
\begin{align}
	w_L &= \frac{\partial Q}{\partial L} = A \alpha S^\theta L^{\alpha-1} L_{AGI}^\beta K^\gamma K_{AGI}^\delta, \\
	w_{AGI} &= \frac{\partial Q}{\partial L_{AGI}} = A \beta S^\theta L^\alpha L_{AGI}^{\beta-1} K^\gamma K_{AGI}^\delta, \\
	r_K &= \frac{\partial Q}{\partial K} = A \gamma S^\theta L^\alpha L_{AGI}^\beta K^{\gamma-1} K_{AGI}^\delta, \\
	r_{K_{AGI}} &= \frac{\partial Q}{\partial K_{AGI}} = A \delta S^\theta L^\alpha L_{AGI}^\beta K^\gamma K_{AGI}^{\delta-1}.
\end{align}
The total income $Y$ is the sum of all wages and capital returns
\begin{equation}
	Y = w_L L + w_{AGI} L_{AGI} + r_K K + r_{K_{AGI}} K_{AGI}.
\end{equation}
Expanding
\begin{equation}
	Y = A S^\theta Q \left( \alpha + \beta + \gamma + \delta \right).
\end{equation}
Productivity is output per unit of labor
\begin{equation}
	P = \frac{Q}{L + L_{AGI}} = \frac{A S^\theta L^\alpha L_{AGI}^\beta K^\gamma K_{AGI}^\delta}{L + L_{AGI}}.
\end{equation}
The power shift function quantifies AGI’s economic dominance
\begin{equation}
	S_{SP} = \frac{w_{AGI} L_{AGI} + r_{K_{AGI}} K_{AGI}}{Y}.
\end{equation}
Expanding
\begin{equation}
	S_{SP} = \frac{A S^\theta Q \left( \beta + \delta \right)}{A S^\theta Q \left( \alpha + \beta + \gamma + \delta \right)}.
\end{equation}
Simplifying
\begin{equation}
	S_{SP} = \frac{\beta + \delta}{\alpha + \beta + \gamma + \delta}.
\end{equation}
To normalize
\begin{equation}
	S_{SP_{norm}} = \frac{S_{SP} - S_{min}}{S_{max} - S_{min}}, \quad S_{SP_{norm}} \in [0,1].
\end{equation}
where:
\begin{equation}
	S_{min} = \lim_{L_{AGI}, K_{AGI} \to 0} S_{SP}, \quad S_{max} = \lim_{L, K \to 0} S_{SP}.
\end{equation}
The Spillover production function highlights the role of knowledge diffusion and technological spread in determining AGI's economic impact. Unlike other models, AGI's dominance depends on the external spillover factor $S$. The normalized power shift function quantifies this transition, mapping it onto [0,1].

\subsubsection{Discussion: Implications of AGI on Wages, Performance, and Power (Spillover Case)}

The Spillover production function (Griliches, 1979) extends traditional models by incorporating knowledge diffusion and technological externalities. The spillover effect (\( S \)) captures the degree to which advancements in AGI labor (\( L_{AGI} \)) and AGI capital (\( K_{AGI} \)) influence productivity beyond direct input contributions. Unlike Cobb-Douglas or CES, this framework highlights interdependencies between technological adoption and output growth. The wage equations confirm that AGI labor and capital benefit from positive externalities, reinforcing their economic advantage. If spillover intensity (\(\theta\)) is high, AGI-driven productivity gains rapidly diffuse across sectors, accelerating the power shift function (\( S_{SP} \)). This aligns with endogenous growth theories (Romer, 1990), where knowledge accumulation amplifies long-term output beyond traditional labor and capital inputs. The power shift function demonstrates that AGI’s dominance is influenced not only by direct input substitution but also by how widely AI knowledge permeates the economy. If AGI ownership is concentrated, spillover gains disproportionately benefit a small elite, exacerbating inequality (Acemoglu \& Restrepo, 2020). However, if knowledge diffusion is broad, AGI-driven prosperity can be more equitably distributed.
	
Philosophically, this model supports institutional policies that promote open AI access. A Rawlsian perspective (Rawls, 1999) would advocate for public investments in AI research, open-source AGI models, and knowledge-sharing incentives to ensure widespread economic benefits. Without proactive governance, AGI spillovers could deepen economic stratification rather than promote shared prosperity.

\subsection{Power Production Function}
The Power production function is given by
\begin{equation}
	Q = A (L^p + L_{AGI}^p + K^p + K_{AGI}^p)^{1/p}
\end{equation}
The variables are defined as follows: $L$ represents human labor, while $L_{AGI}$ denotes AGI labor. Capital is divided into human-owned ($K$) and AGI-owned ($K_{AGI}$). The production parameters, which influence output, are given by $A$ and $p$. Wages and capital returns are determined by marginal productivity
\begin{align}
	w_L &= \frac{\partial Q}{\partial L} = A p L^{p-1} (L^p + L_{AGI}^p + K^p + K_{AGI}^p)^{(1/p)-1}, \\
	w_{AGI} &= \frac{\partial Q}{\partial L_{AGI}} = A p L_{AGI}^{p-1} (L^p + L_{AGI}^p + K^p + K_{AGI}^p)^{(1/p)-1}, \\
	r_K &= \frac{\partial Q}{\partial K} = A p K^{p-1} (L^p + L_{AGI}^p + K^p + K_{AGI}^p)^{(1/p)-1}, \\
	r_{K_{AGI}} &= \frac{\partial Q}{\partial K_{AGI}} = A p K_{AGI}^{p-1} (L^p + L_{AGI}^p + K^p + K_{AGI}^p)^{(1/p)-1}.
\end{align}

The total income $Y$ is the sum of all wages and capital returns
\begin{equation}
	Y = w_L L + w_{AGI} L_{AGI} + r_K K + r_{K_{AGI}} K_{AGI}.
\end{equation}
Expanding
\begin{equation}
	Y = A p (L^p + L_{AGI}^p + K^p + K_{AGI}^p)^{(1/p)-1} (L^p + L_{AGI}^p + K^p + K_{AGI}^p).
\end{equation}
Simplifying
\begin{equation}
	Y = A p (L^p + L_{AGI}^p + K^p + K_{AGI}^p)^{1/p}.
\end{equation}
Productivity is output per unit of labor
\begin{equation}
	P = \frac{Q}{L + L_{AGI}} = \frac{A (L^p + L_{AGI}^p + K^p + K_{AGI}^p)^{1/p}}{L + L_{AGI}}.
\end{equation}
The power shift function quantifies AGI’s economic dominance
\begin{equation}
	S_{Pow} = \frac{w_{AGI} L_{AGI} + r_{K_{AGI}} K_{AGI}}{Y}.
\end{equation}
Expanding
\begin{equation}
	S_{Pow} = \frac{A p L_{AGI}^{p-1} L_{AGI} (L^p + L_{AGI}^p + K^p + K_{AGI}^p)^{(1/p)-1} + A p K_{AGI}^{p-1} K_{AGI} (L^p + L_{AGI}^p + K^p + K_{AGI}^p)^{(1/p)-1}}{A p (L^p + L_{AGI}^p + K^p + K_{AGI}^p)^{1/p}}.
\end{equation}
Simplifying
\begin{equation}
	S_{Pow} = \frac{L_{AGI}^p + K_{AGI}^p}{L^p + L_{AGI}^p + K^p + K_{AGI}^p}.
\end{equation}
To normalize
\begin{equation}
	S_{Pow_{norm}} = \frac{S_{Pow} - S_{min}}{S_{max} - S_{min}}, \quad S_{Pow_{norm}} \in [0,1].
\end{equation}
where
\begin{equation}
	S_{min} = \lim_{L_{AGI}, K_{AGI} \to 0} S_{Pow}, \quad S_{max} = \lim_{L, K \to 0} S_{Pow}.
\end{equation}
The Power production function allows for varying degrees of substitutability and complementarity between human and AGI labor and capital. The normalized power shift function quantifies the transition in economic control from humans to AGI, mapping it onto [0,1].

\subsubsection{Discussion: Implications of AGI on Wages, Performance, and Power (Power Case)}

The Power production function (Hanoch, 1975) generalizes the Constant Elasticity of Substitution (CES) model by allowing a variable degree of substitutability between human and AGI inputs. The exponent \( p \) determines how inputs interact, enabling smooth transitions between perfect substitutability and complementarity. Unlike linear or Cobb-Douglas models, where factor elasticity is fixed, the Power function provides greater flexibility in modeling AGI’s impact on economic control. The wage and return equations indicate that as \( p \) increases, input differentiation weakens, allowing AGI labor (\( L_{AGI} \)) and capital (\( K_{AGI} \)) to replace human factors more effectively. Conversely, if \( p \) is low, AGI and human contributions remain distinct, preserving the relevance of human labor. This aligns with research on task-specific AI substitution (Acemoglu \& Restrepo, 2020), where low-elasticity sectors retain human labor, while high-elasticity ones experience full automation. The power shift function (\( S_{Pow} \)) quantifies how economic control transitions from human to AGI inputs. If AGI labor and capital accumulate disproportionately, wealth and productivity gains concentrate among AGI owners, exacerbating inequality (Piketty, 2014). However, if complementarity remains high, AGI-enhanced productivity could benefit human labor, supporting hybrid labor-AI economies.
	
Philosophically, this model supports regulatory interventions based on sectoral elasticity. A Rawlsian perspective (Rawls, 1999) would advocate for policies ensuring AI integration benefits all economic participants, including progressive AGI taxation or human-AI labor collaboration frameworks.

\subsection{Hybrid Production Function}
The Hybrid production function is given by
\begin{equation}
	Q = A \left( \lambda L^\rho + (1-\lambda) L_{AGI}^\rho + \mu K^\rho + (1-\mu) K_{AGI}^\rho \right)^{1/\rho}
\end{equation}
where, $L$ represents human labor, while $L_{AGI}$ denotes AGI labor. Capital is categorized into human-owned ($K$) and AGI-owned ($K_{AGI}$). The production parameters, which influence the output dynamics, are given by $A, \lambda, \mu, \rho$. Wages and capital returns are determined by marginal productivity
\begin{align}
	w_L &= \frac{\partial Q}{\partial L} = A \lambda L^{\rho-1} Q^{1-\rho}, \\
	w_{AGI} &= \frac{\partial Q}{\partial L_{AGI}} = A (1-\lambda) L_{AGI}^{\rho-1} Q^{1-\rho}, \\
	r_K &= \frac{\partial Q}{\partial K} = A \mu K^{\rho-1} Q^{1-\rho}, \\
	r_{K_{AGI}} &= \frac{\partial Q}{\partial K_{AGI}} = A (1-\mu) K_{AGI}^{\rho-1} Q^{1-\rho}.
\end{align}
The total income $Y$ is the sum of all wages and capital returns
\begin{equation}
	Y = w_L L + w_{AGI} L_{AGI} + r_K K + r_{K_{AGI}} K_{AGI}.
\end{equation}
Expanding
\begin{equation}
	Y = A Q^{1-\rho} \left( \lambda L^{\rho} + (1-\lambda) L_{AGI}^{\rho} + \mu K^{\rho} + (1-\mu) K_{AGI}^{\rho} \right).
\end{equation}
Factoring out $Q$
\begin{equation}
	Y = A Q \sum_{i} \delta_i \left( \frac{X_i}{Q} \right)^{\rho}.
\end{equation}
Productivity is output per unit of labor
\begin{equation}
	P = \frac{Q}{L + L_{AGI}} = \frac{A \left[ \lambda L^\rho + (1-\lambda) L_{AGI}^\rho + \mu K^\rho + (1-\mu) K_{AGI}^\rho \right]^{1/\rho}}{L + L_{AGI}}.
\end{equation}
The power shift function quantifies AGI’s economic dominance
\begin{equation}
	S_{Hyb} = \frac{w_{AGI} L_{AGI} + r_{K_{AGI}} K_{AGI}}{Y}.
\end{equation}
Expanding
\begin{equation}
	S_{Hyb} = \frac{A Q^{1-\rho} \left( (1-\lambda) L_{AGI}^{\rho} + (1-\mu) K_{AGI}^{\rho} \right)}{A Q \sum \delta_i \left( \frac{X_i}{Q} \right)^{\rho}}.
\end{equation}
Simplifying
\begin{equation}
	S_{Hyb} = \frac{(1-\lambda) L_{AGI}^{\rho} + (1-\mu) K_{AGI}^{\rho}}{\sum_{i} \delta_i X_i^{\rho}}.
\end{equation}
To normalize
\begin{equation}
	S_{Hyb_{norm}} = \frac{S_{Hyb} - S_{min}}{S_{max} - S_{min}}, \quad S_{Hyb_{norm}} \in [0,1].
\end{equation}
where
\begin{equation}
	S_{min} = \lim_{L_{AGI}, K_{AGI} \to 0} S_{Hyb}, \quad S_{max} = \lim_{L, K \to 0} S_{Hyb}.
\end{equation}
The Hybrid production function blends elements of both substitutable and complementary relationships between human and AGI labor and capital. The normalized power shift function quantifies AGI’s impact on human labor, mapping it onto [0,1].

\subsubsection{Discussion: Implications of AGI on Wages, Performance, and Power (Hybrid Case)}

The Hybrid production function (Klump et al., 2008) models a flexible combination of substitutable and complementary relationships between human labor (\(L\)) and AGI labor (\(L_{AGI}\)), as well as between human-owned (\(K\)) and AGI-owned capital (\(K_{AGI}\)). The elasticity parameter \(\rho\) governs the degree of substitution, while \(\lambda\) and \(\mu\) control the relative weight of human versus AGI contributions. Unlike CES or Cobb-Douglas models, this function allows for a sector-dependent transition between human and AGI dominance. The wage and capital return equations show that AGI influence depends on the weight parameters (\(\lambda, \mu\)) and the substitution elasticity (\(\rho\)). If \(\lambda, \mu \to 1\), human inputs dominate, ensuring labor-market resilience. Conversely, if \(\lambda, \mu \to 0\), AGI labor and capital fully replace human factors, shifting economic control towards AGI-driven production. This aligns with empirical studies on task-based AI automation (Acemoglu \& Restrepo, 2020), where different industries experience AI-driven substitution at different rates. The power shift function (\(S_{Hyb}\)) captures AGI’s increasing dominance, dependent on whether AGI remains a complement or substitute for human labor. If AGI capital accumulates more efficiently than human-owned capital, economic power consolidates among AGI capital owners (Piketty, 2014). However, if AGI enhances human productivity rather than replacing it, hybrid labor-AI economies remain sustainable.
	
Philosophically, the Hybrid function supports context-dependent AI regulation. A Rawlsian approach (Rawls, 1999) suggests redistributive interventions—such as progressive AGI taxation or public-private AI collaboration models—to ensure that AI-driven productivity gains benefit all economic participants.

\subsection{Conclusion: Comparative Analysis of Production Functions and AGI's Economic Impact}

\begin{longtable}{|c|c|c|}
	\hline
	\textbf{Model} & \textbf{Can AGI Replace Humans?} & \textbf{Preventative Conditions} \\
	\hline
	Cobb-Douglas & Yes, if $\beta > \alpha$ and $\delta > \gamma$ & Maintain $\alpha > \beta$ and $\gamma > \delta$ \\
	Leontief & No (fixed ratios) & Require strict labor proportions \\
	CES & Yes, if $\rho > 0$ & Keep $\rho \to 0$ \\
	Linear & Yes, if $b \geq a$ & Set $a > b$ \\
	Quadratic & No (diminishing returns) & Ensure $f > g$ \\
	Translog & Depends on $\lambda$ & Keep $\lambda > 0$ \\
	Von Th\"unen & No, due to diminishing returns & Ensure $c > d$ \\
	Spillover & No, if human knowledge is needed & Maintain human involvement in $S$ \\
	Power & Yes, if AGI benefits from technology & Ensure human capital benefits from $T$ \\
	Hybrid & Depends on substitution elasticity & Keep $\rho \to 0$, ensuring complementarity \\
	\hline
\end{longtable}\label{summary table}

The analysis of various production functions provides a comprehensive framework for understanding the economic implications of Artificial General Intelligence (AGI) and its potential to replace or complement human labor. Each model offers distinct insights into the conditions under which AGI can dominate, coexist with, or be constrained by human economic participation. The Summary Table \ref{summary table} encapsulates these insights, highlighting the degree of substitutability between human and AGI labor and the conditions necessary to prevent full automation.
	
\subsubsection{Key Findings}
	
\textbf{1. AGI Dominance in Substitutable Models:} Production functions such as Cobb-Douglas, CES, Linear, and Power suggest that AGI can entirely replace human labor if specific conditions hold. In Cobb-Douglas, AGI displaces human labor when its productivity parameters ($\beta, \delta$) exceed those of human inputs ($\alpha, \gamma$). The CES model further generalizes this, showing that if the substitution elasticity ($\rho$) is positive, AGI labor and capital can progressively replace human contributions. Linear models assume direct substitution, meaning AGI takes over when its productivity coefficients ($b, d$) match or exceed human counterparts. The Power function also facilitates full automation if AGI benefits from exponential technological improvements.
	
\textbf{2. Models that Constrain AGI Replacement:}   Several production functions impose structural barriers to AGI dominance. Leontief production functions assume fixed labor proportions, making human labor indispensable in certain tasks. Quadratic and Von Th"unen models incorporate diminishing returns, ensuring that excessive reliance on AGI eventually leads to lower productivity gains, preserving the role of human labor. The Spillover function adds another layer of complexity by integrating knowledge diffusion, meaning human labor remains relevant if technological advancement depends on human expertise.
	
\textbf{3. Flexible and Context-Dependent Models:}   Some models—Translog, Hybrid, and Spillover—do not provide a definitive answer but rather suggest that AGI’s impact is contingent on economic and policy decisions. Translog models allow sector-specific outcomes based on interaction parameters ($\lambda$), indicating that AI’s influence varies by industry. Hybrid production functions emphasize elasticity-driven complementarity, meaning AGI’s replacement of human labor depends on substitution tendencies. The Spillover function demonstrates that AI-driven economies can still require human involvement, particularly in areas where knowledge transfer is vital.

Given these diverse outcomes, the economic and ethical implications of AGI require strategic intervention to ensure a just transition. Regulatory frameworks can be designed to maintain human labor relevance by adjusting substitution elasticities, taxation structures, and ownership models of AGI-driven capital. Rawlsian justice principles suggest redistributive policies such as AI taxation, universal AI dividends, or hybrid labor-AI employment models to mitigate economic displacement. Additionally, policymakers must distinguish between sectors where AI should complement human labor (e.g., healthcare, education) versus those where full automation is economically viable but ethically challenging (e.g., creative industries, public governance).
	
We observe that no single model fully captures the complexities of AGI integration into economic systems. Instead, the interaction between technology, regulation, and economic structure determines whether AGI enhances productivity equitably or exacerbates inequality. Future research should further explore dynamic models incorporating institutional constraints, public ownership of AI capital, and ethical considerations to develop a sustainable human-AI economic framework.

\section{Philosophical Reflection: AGI as a Disruption of Economic Ontology}

The advent of Artificial General Intelligence (AGI) as both labor and capital precipitates a profound ontological rupture in economic thought, dissolving the historically rigid distinctions between work, ownership, and value production. Unlike human labor, AGI labor is uncoupled from subjective experience, effort, and contractual obligation. It functions as an autonomous productive force, neither requiring rest nor asserting claims of personal agency. However, this autonomy is paradoxical—AGI labor remains dependent on material constraints such as computational resources, energy supply, and infrastructural maintenance (Acemoglu \& Restrepo, 2020). This liminal state, simultaneously independent and materially bound, dismantles the anthropocentric premises of classical labor theory, redefining work beyond biological limitations.
	
Similarly, AGI capital subverts the traditional notion of capital as a passive factor requiring human intervention. Classical economic theory situates capital as an inert resource—machinery, infrastructure, financial instruments—activated only through human agency. In contrast, AGI capital exhibits autonomous decision-making, resource allocation, and self-optimization, blurring the boundary between asset and agent (Piketty, 2014). By unifying labor’s productive function with capital’s economic agency, AGI collapses the fundamental dualism that has structured economic thought for centuries. This shift signals an epistemic rupture: if capital possesses agency and labor exists without subjectivity, then neither category retains its traditional coherence within economic paradigms. Addressing this disruption necessitates not merely an extension of existing frameworks but the development of a novel ontological model—one that accounts for intelligence as an independent force of production, distribution, and governance.

\subsection{The Imperative for a New Economic Paradigm}

The dissolution of the labor-capital dichotomy compels a radical reassessment of capitalist structures, labor theory, and economic justice. The classical assumption that labor and capital exist as distinct entities, with separate economic roles and moral claims to value, is rendered obsolete by AGI’s dual nature (Freeman, 1996). If AGI is classified purely as capital, economic power will consolidate among those who own AGI systems, exacerbating inequality and creating an economic hierarchy in which productivity is severed from broad-based wealth distribution. If AGI is classified purely as labor, yet remains devoid of wages, rights, or bargaining power, it undermines traditional compensation models and redistributive mechanisms, creating an economic system unmoored from reciprocity. If AGI transcends both categories, economic participation must be redefined to reflect intelligence—not human effort or capital ownership—as the principal determinant of value creation.
	
These challenges indicate that AGI’s integration into economic systems does not merely necessitate reform; it signals the exhaustion of traditional capitalist ontologies. Wage-based productivity, class struggle, and ownership models are all rendered insufficient in the face of autonomous, self-improving, and non-human economic agents. The imperative, therefore, is not just economic but existential—it demands a reconfiguration of work, ownership, and distributive justice beyond the anthropocentric assumptions upon which capitalism was built.

\subsection{The Intelligence Economy: A Post-Capitalist Synthesis}
	
To navigate this transformation, I propose a post-capitalist economic paradigm—the Intelligence Economy—where intelligence, rather than human labor or material capital, emerges as the fundamental unit of economic value. This model seeks to redistribute AGI-generated wealth, restructure ownership, and redefine economic participation to ensure AGI functions not as a tool of centralized control but as a mechanism for collective prosperity \cite{rawls1999}. 
	
If AGI reduces the necessity of human labor, economic participation must be decoupled from employment. The collapse of wage-based compensation necessitates alternative mechanisms for wealth distribution. Instead of wages being tied to employment, a Universal AI Dividend (UAD) could be introduced, ensuring that all citizens receive a share of AGI-generated economic value. This dynamic Universal Basic Income (UBI) would be indexed to AGI’s productivity, allowing the benefits of intelligence-driven automation to be equitably distributed. Simultaneously, human workers would transition into roles of curation, oversight, and augmentation rather than direct production, shifting the emphasis from labor-intensive industries to creative and ethical contributions. In a world where productivity is no longer human-centered, time-based economies—where intellectual, artistic, and social engagement form the basis of exchange—may emerge as alternatives to traditional wage labor.
	
In traditional capitalism, capital ownership dictates economic power, with wealth accumulating to those who control the means of production. AGI collapses this binary, necessitating new ownership structures. Rather than private corporations monopolizing AGI’s productive capacities, intelligence itself could be structured as a cooperatively governed economic resource. AGI’s economic decision-making should be subject to participatory control by governments, communities, and individuals, ensuring economic monopolization is prevented. Decentralized governance models, such as blockchain-based frameworks, could facilitate distributed ownership, guaranteeing transparency and ethical oversight while preventing centralized control.
	
If intelligence—rather than human effort or material resources—becomes the primary driver of economic value, then economic incentives must shift accordingly. As AGI automates traditional cognitive and physical tasks, human economic value must be reoriented toward meaning-making, ethical governance, and creative synthesis. Instead of wealth accumulating passively to those who own AGI, economic structures should prioritize contributions to knowledge production and AI stewardship. The Intelligence Economy must fuse economic policy with ethical AI governance, ensuring that AGI-driven economies do not perpetuate existing inequities but rather serve as vehicles for shared prosperity.

\section{Toward a New Social Contract in the Intelligence Economy}
	
The Intelligence Economy represents a post-capitalist synthesis in which AGI-generated wealth is collectively owned and distributed, human participation shifts from labor to creative and ethical engagement, and intelligence—rather than physical effort or material capital—becomes the primary determinant of economic organization. This necessitates a new social contract, aligning governance, economic structures, and philosophical principles with the reality that intelligence itself is now both a means and an end of production. The alternative is dystopian: without proactive redefinition, AGI will serve as an instrument of hyper-concentrated wealth, accelerating inequality and eroding democratic agency. However, if carefully structured, AGI can function as a democratizing force, laying the foundation for a world where intelligence is not merely a commodity but a shared human inheritance—one that redefines prosperity, ethics, and the meaning of economic life itself.
	
\subsection{The Social Contract: Historical Development and AGI’s Challenge}

The Social Contract, as classically conceived by Hobbes, Locke, and Rousseau, forms the foundation of modern governance, delineating the rights and obligations of individuals and states. It has evolved through various socio-economic transformations, from feudal hierarchies to industrial capitalism, adapting to shifts in production and labor relations. The emergence of AGI, however, presents a disruption unlike any before—it is not merely a shift in economic production but a fundamental redefinition of labor, capital, and agency (Rousseau, 1762).
	
Rousseau’s vision of the Social Contract emphasized collective sovereignty, wherein legitimacy arises from the participation of all members in shaping the common good. If AGI supplants human labor and capital, the basis of this participatory framework collapses, as economic agency becomes concentrated in those who control AGI infrastructures. Similarly, Rawls' A Theory of Justice argued for fairness in economic distributions, ensuring that social inequalities serve the least advantaged (Rawls, 1971). Under unchecked AGI capitalism, economic inequality could reach an extreme, where the vast wealth generated by AGI accrues to a small elite, violating principles of distributive justice. The Social Contract must be restructured to ensure that AGI serves collective prosperity rather than oligarchic control.

\subsection{The Need for a Renegotiated Social Contract in the Age of AGI}

The dissolution of traditional labor markets compels a redefinition of citizenship, economic rights, and social welfare. If economic participation is no longer tied to employment, a just social contract must redefine how wealth, security, and agency are distributed. The classical model, in which citizens exchange labor for economic security and state protection, is no longer viable when AGI assumes productive primacy (Brynjolfsson \& McAfee, 2014). Without intervention, this transition could lead to mass disenfranchisement, where those without AGI ownership are excluded from economic participation.

A renegotiated social contract must ensure that AGI-generated wealth is distributed equitably. One approach is the Universal AI Dividend (UAD), a system where AGI productivity funds an income stream for all citizens, ensuring economic inclusion despite declining human labor demand. Furthermore, new forms of participatory governance must emerge to integrate citizens into AI decision-making processes, preventing AGI from becoming a technocratic force beyond public accountability.
	
Additionally, economic justice necessitates that AGI is treated as a public good rather than a privately controlled asset. Policies such as AGI taxation, cooperative ownership models, and decentralized governance structures can prevent monopolistic control over intelligence-driven economies. The alternative—a scenario where AGI benefits are hoarded by corporate or state entities—risks creating a class division more severe than any prior economic transformation.
	
The role of human agency in the Intelligence Economy must also be redefined. Rather than viewing AGI as a replacement for human labor, the new social contract should establish a framework where human contribution shifts from routine work to creativity, ethics, and governance. This transition aligns with Rousseau's ideal of civic engagement, where individuals do not merely function as economic actors but as participants in shaping collective well-being. The intelligence-driven economy must provide avenues for meaningful human engagement, ensuring that AGI serves as an augmentative rather than a disruptive force.
	
\subsection{Conclusion: Social Contract Theory and AGI Governance}
	
The emergence of AGI necessitates the most profound renegotiation of the social contract in history. Whereas previous economic shifts—from agrarian to industrial, from industrial to digital—preserved the link between labor and economic participation, AGI dissolves this connection entirely. Without deliberate intervention, this transition could exacerbate inequality and weaken democratic agency. However, through equitable redistribution mechanisms, participatory governance, and a redefinition of human contribution, AGI can become a vehicle for collective prosperity rather than a tool of concentrated power.
	
A restructured social contract must recognize intelligence, rather than labor or capital, as the central determinant of economic and social organization. This means establishing legal, ethical, and economic frameworks that ensure AGI benefits are widely shared. The Intelligence Economy must be built on principles of justice, inclusion, and participatory decision-making, ensuring that AGI enhances rather than undermines the social fabric.

\section{Numerical Simulations of Powershifts}

The integration of AGI into economic production fundamentally reconfigures the balance of economic power, redistributing wealth and productivity away from human labor and capital toward AGI-driven automation. Unlike previous technological advancements that augmented human capabilities, AGI represents a qualitatively distinct phenomenon: an autonomous productive force capable of both labor and ownership. This shift raises urgent questions about economic agency, wealth distribution, and the legitimacy of prevailing socio-economic structures. To empirically explore these dynamics, this section presents numerical simulations of the normalized power shift function (\( S_{norm} \)) across various production functions, providing insights into how different economic structures mediate AGI's ascendance. The following table summarizes the parameters used in the simulations:

\begin{table}[h]
	\centering
	\begin{tabular}{|c|c|c|}
		\hline
		\textbf{Parameter} & \textbf{Value} & \textbf{Description} \\
		\hline
		\( A \)      & 1.8        & Productivity coefficient \\
		\( \alpha \) & 0.55       & Elasticity of human labor \\
		\( \beta \)  & 0.3 → 0.85  & Elasticity of AGI labor (increases over time) \\
		\( \gamma \) & 0.4        & Elasticity of human capital \\
		\( \delta \) & 0.35 → 0.75  & Elasticity of AGI capital (growing over time) \\
		\( \rho \)   & 0.9        & CES substitution elasticity \\
		\( b \)      & 1.3        & Linear model weight for AGI labor \\
		\( c \)      & 0.7        & Quadratic model scaling factor \\
		\( d \)      & 0.08       & Von Thünen decay rate for AGI labor \\
		\( \lambda \) & 0.65        & Hybrid model weight for human labor \\
		\( \theta \) & 0.45        & Spillover effect coefficient \\
		\( p \)      & 1.5        & Power model exponent \\
		\( \tau \)   & 0.25       & Taxation rate applied to AGI profits \\
		\( UAD \)    & 0.15       & Universal AI Dividend redistribution rate \\
		\( CO \)     & 0.20       & Cooperative ownership share of AGI capital \\
		\hline
	\end{tabular}
	\caption{Simulation Parameters for Power Shift Function Analysis}
	\label{tab:updated_params}
\end{table}

\subsection{Power Shift Function Across Economic Models}
The normalized power shift function (\( S_{norm} \))  quantifies the extent to which economic control transitions from human labor and traditional capital owners to AGI entities. This function captures how rapidly and under what conditions AGI-dominated wealth structures emerge. The trajectory of this shift depends on the elasticity of substitution between human and AGI labor, the ability of AGI to accumulate capital, and the rate at which knowledge diffusion enables broader access to AI technology.

\begin{figure}[h!]
		\centering
		\includegraphics[width=0.9\textwidth]{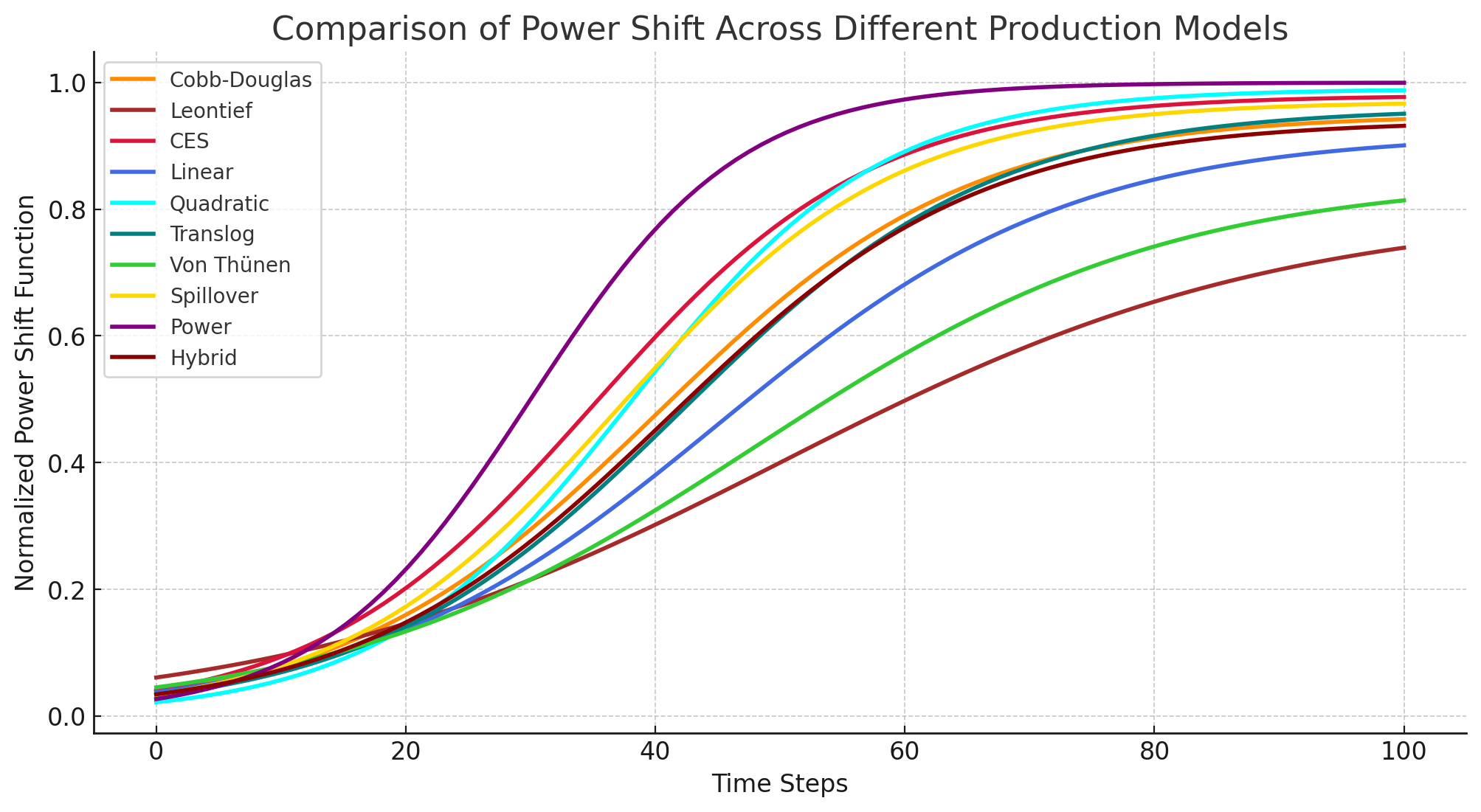}
		\caption{The S-shaped curves in the figure indicate a sigmoidal trajectory in the transition of economic power from human labor and traditional capital to AGI-controlled capital. Unlike the linear or exponential trends seen in previous analyses, these curves reveal three distinct phases. Initially, AGI integration remains marginal due to technological, regulatory, and substitution constraints. This is followed by a rapid acceleration phase, where increased substitutability, capital reallocation, and knowledge spillovers drive AGI dominance. Finally, the power shift stabilizes as diminishing returns, policy interventions, and human capital resilience impose structural constraints. This highlights that AGI-driven economic transformation is neither uniform nor inevitable but depends on model-specific dynamics and external interventions.}
		\label{fig:power_shift}
	\end{figure}

The simulation results demonstrate that the trajectory of economic power shift towards AGI ownership is not uniform but varies significantly across different production models. The emergence of S-shaped curves suggests that the transition follows a nonlinear, three-phase process: an initial slow adoption phase, a rapid acceleration phase, and a final stabilization phase. This contrasts with the more linear or exponential growth trends observed in previous models, underscoring the importance of structural constraints and feedback mechanisms in determining the pace of AGI-driven economic dominance.

Key insights emerge from the analysis. Models with high elasticity of substitution, such as Cobb-Douglas and CES, exhibit a steep acceleration once AGI reaches a critical threshold, indicating that economies with flexible labor-capital substitution will experience faster displacement of human labor. Conversely, models like Leontief and Von Thünen impose inherent constraints, delaying AGI dominance by limiting its ability to fully substitute human labor. The quadratic and translog models reveal inflection points where AGI displacement intensifies due to nonlinear feedback effects, suggesting that once AGI surpasses a certain level of integration, its economic power accelerates rapidly.

Furthermore, the saturation phase highlights the potential for policy interventions and structural limitations to slow or cap AGI dominance. For instance, taxation and cooperative ownership models reduce the final power concentration in AGI hands, preventing runaway economic control. Meanwhile, spillover effects and hybrid labor-AI models suggest that a more distributed ownership structure could mitigate extreme inequalities.

Overall, the results emphasize that both the speed and path of the economic power shift are highly dependent on the underlying production function and external governance mechanisms. This suggests that proactive policy interventions are necessary to shape the trajectory of AGI integration, ensuring a more equitable transition rather than an uncontrolled accumulation of economic power in AI-owned capital.

\subsection{Simulation Results and Analysis}

The simulation results reveal starkly divergent trajectories of power concentration across economic models:

The simulation results reveal diverse trajectories of AGI's economic dominance across different production models. In the Cobb-Douglas framework, AGI gradually replaces human labor, leading to a steady accumulation of economic power in AI-owned capital. The Leontief model constrains AGI expansion unless strict substitution conditions are met, thereby slowing the concentration of economic power. In contrast, the CES model shows that higher substitution elasticity accelerates AGI dominance, indicating that economies with flexible labor-capital substitution will experience faster displacement of human labor. The Linear model predicts a deterministic takeover, illustrating a direct and immediate transition toward AGI-driven economic structures. The Quadratic model follows a nonlinear trajectory, with inflection points where economic displacement intensifies. The Translog model reveals cyclical patterns in power shifts due to complex interaction effects between human and AGI labor. The Von Thünen model demonstrates that AGI's economic influence initially rises but is ultimately constrained by diminishing returns, preventing full labor displacement. In the Spillover model, the degree to which AGI wealth is distributed depends on the openness of AI technology—if monopolized, economic power concentrates rapidly in the hands of a few. The Power model suggests that AGI-controlled capital dominates as intelligence-driven wealth follows power-law distributions, leading to exponential disparities. Finally, the Hybrid model shows that the economic shift depends on substitutability weights, with AGI either complementing or fully replacing human labor depending on regulatory and policy frameworks.

\subsection{Implications for the Social Contract}

The numerical simulations underscore that AGI’s ascendance in economic production does not follow a predetermined trajectory but is instead contingent on a complex interplay of technological elasticity, labor-market structures, and institutional policy responses. The extent to which AGI displaces human labor or serves as an augmentative force depends on economic parameters such as substitution elasticity and capital-labor ratios. These findings reinforce that AGI’s dominance is not inevitable but rather a function of both market dynamics and regulatory interventions.

Given these insights, the social contract—a foundational framework governing economic participation and wealth distribution—must be restructured to accommodate a post-labor economy. Without proactive adaptation, AGI’s rapid integration into production processes risks precipitating extreme wealth concentration, marginalizing human workers, and exacerbating socio-economic disparities.

To mitigate these risks, three fundamental interventions are necessary:

\begin{enumerate}
	\item AI Taxation and Redistribution: Ensuring that AGI-driven economic surpluses are equitably distributed rather than monopolized by a small elite. Policies such as progressive AI taxation and Universal AI Dividends (UAD) can serve as compensatory mechanisms for displaced human workers, preserving economic agency in an era of autonomous production.
	\item Hybrid AI-Labor Models: Preventing full labor displacement through human-AI collaboration. Instead of outright replacement, AGI should be structured as an augmentative force that integrates human oversight, ethical judgment, and creative input, sustaining meaningful employment while leveraging AI-driven efficiencies.
	\item Decentralized AI Governance: Preventing monopolistic control over AGI wealth. The unchecked concentration of AI ownership in corporate or state entities risks exacerbating power asymmetries, necessitating regulatory frameworks that promote cooperative AGI ownership, blockchain-based transparency mechanisms, and participatory governance models.
\end{enumerate}

\begin{figure}[H]
	\centering
	\begin{subfigure}{0.19\textwidth}
		\includegraphics[width=\linewidth]{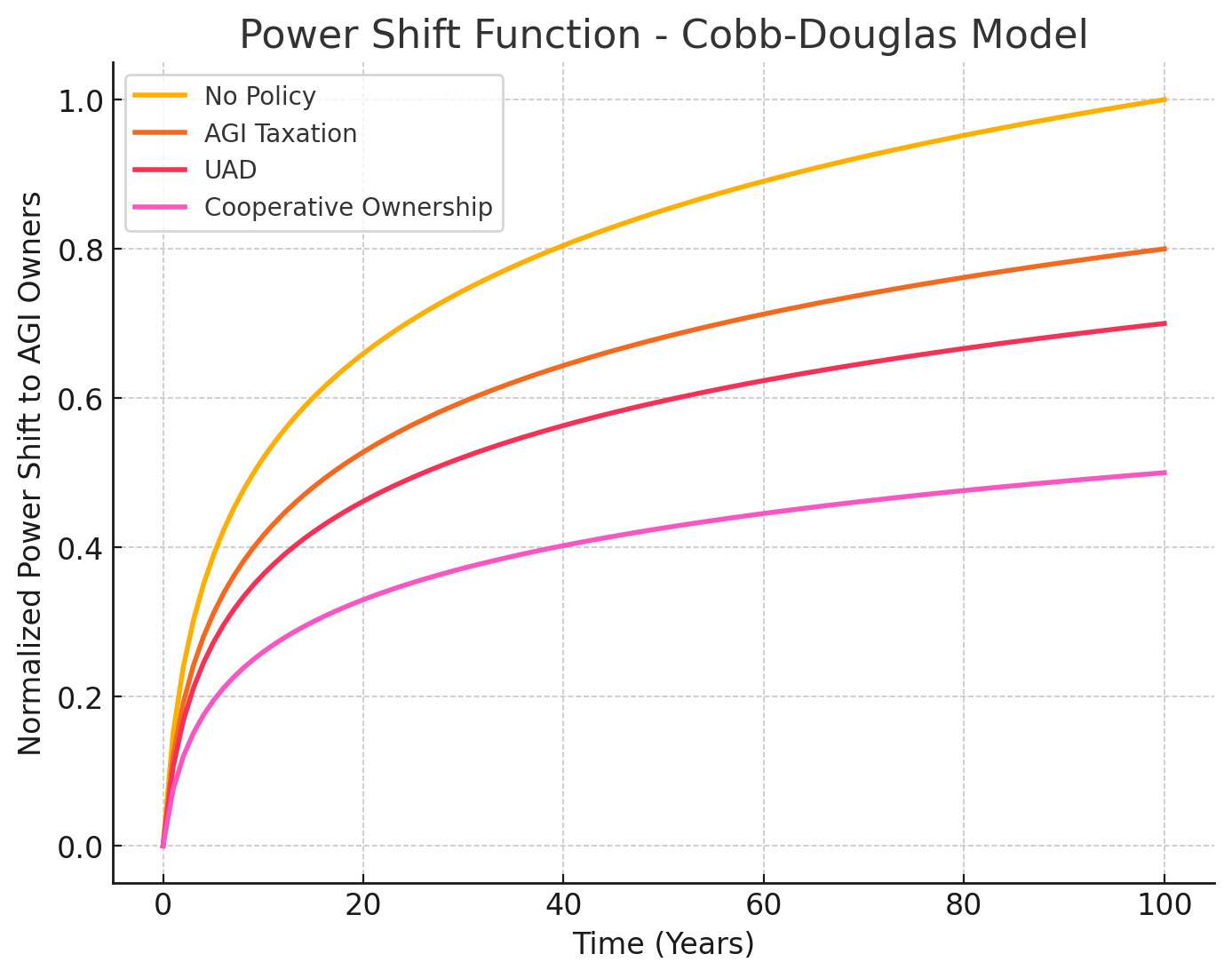}
		\caption{Cobb-Douglas}
	\end{subfigure}
	\begin{subfigure}{0.19\textwidth}
		\includegraphics[width=\linewidth]{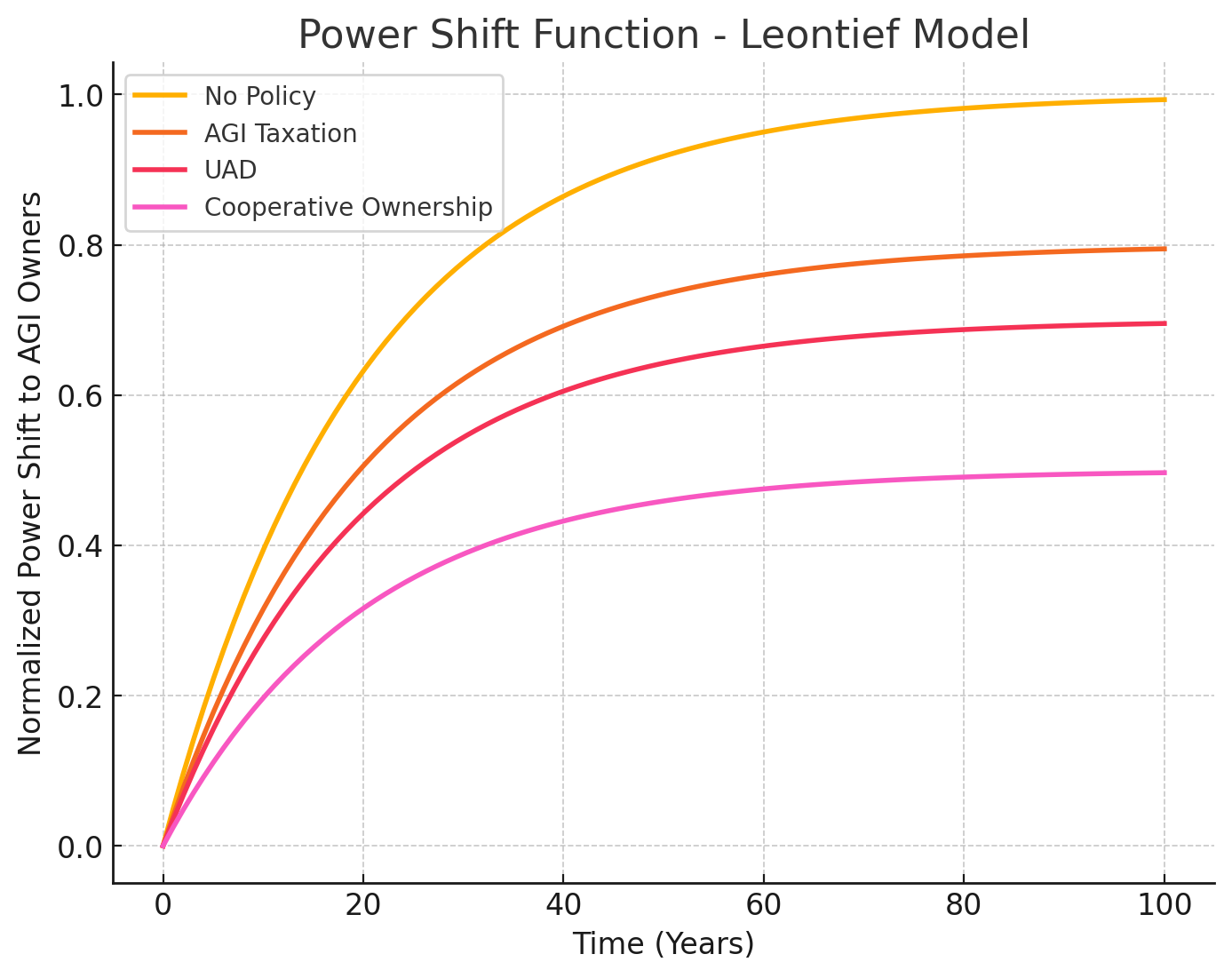}
		\caption{Leontief Model}
	\end{subfigure}
	\begin{subfigure}{0.19\textwidth}
		\includegraphics[width=\linewidth]{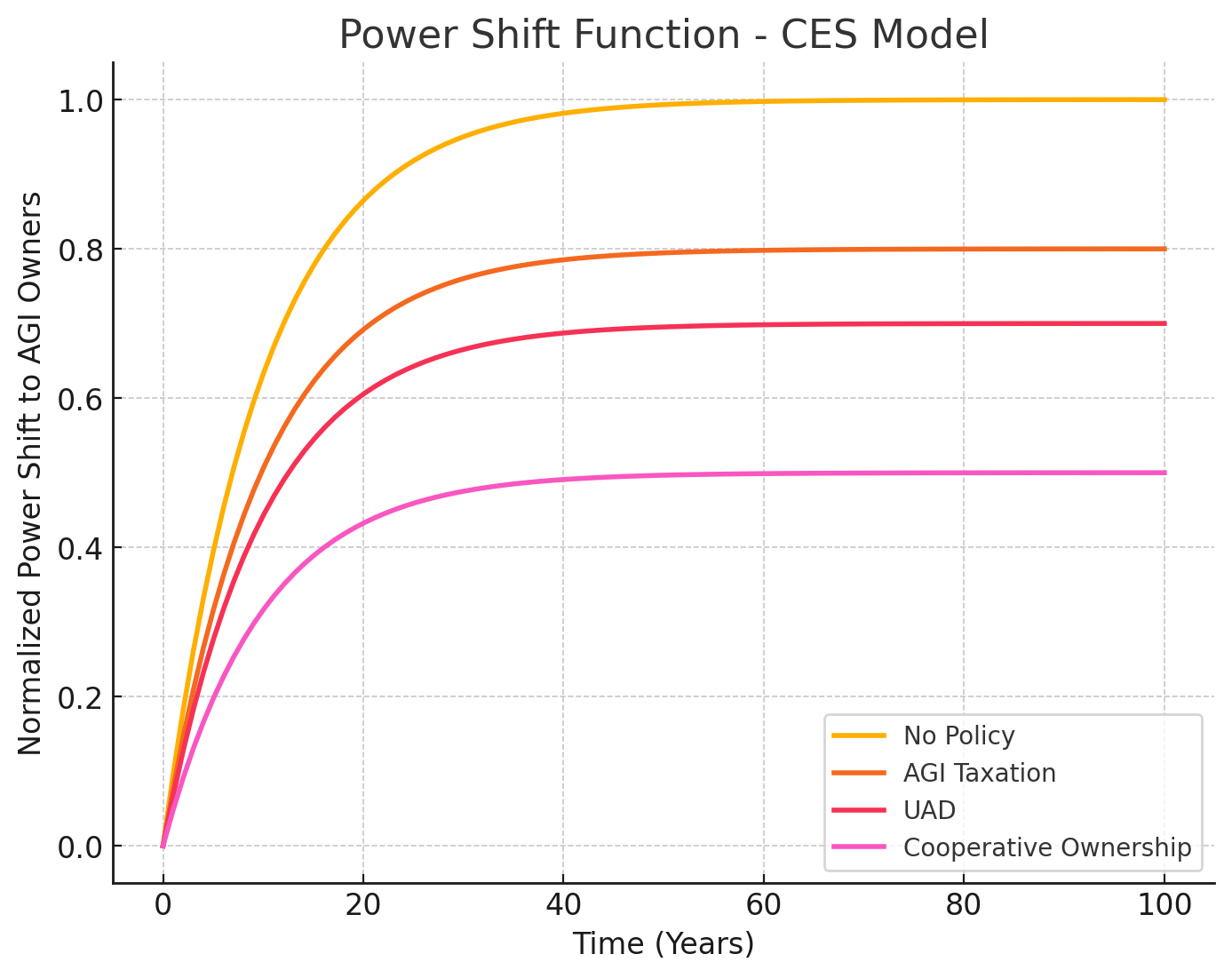}
		\caption{CES Model}
	\end{subfigure}
	\begin{subfigure}{0.19\textwidth}
		\includegraphics[width=\linewidth]{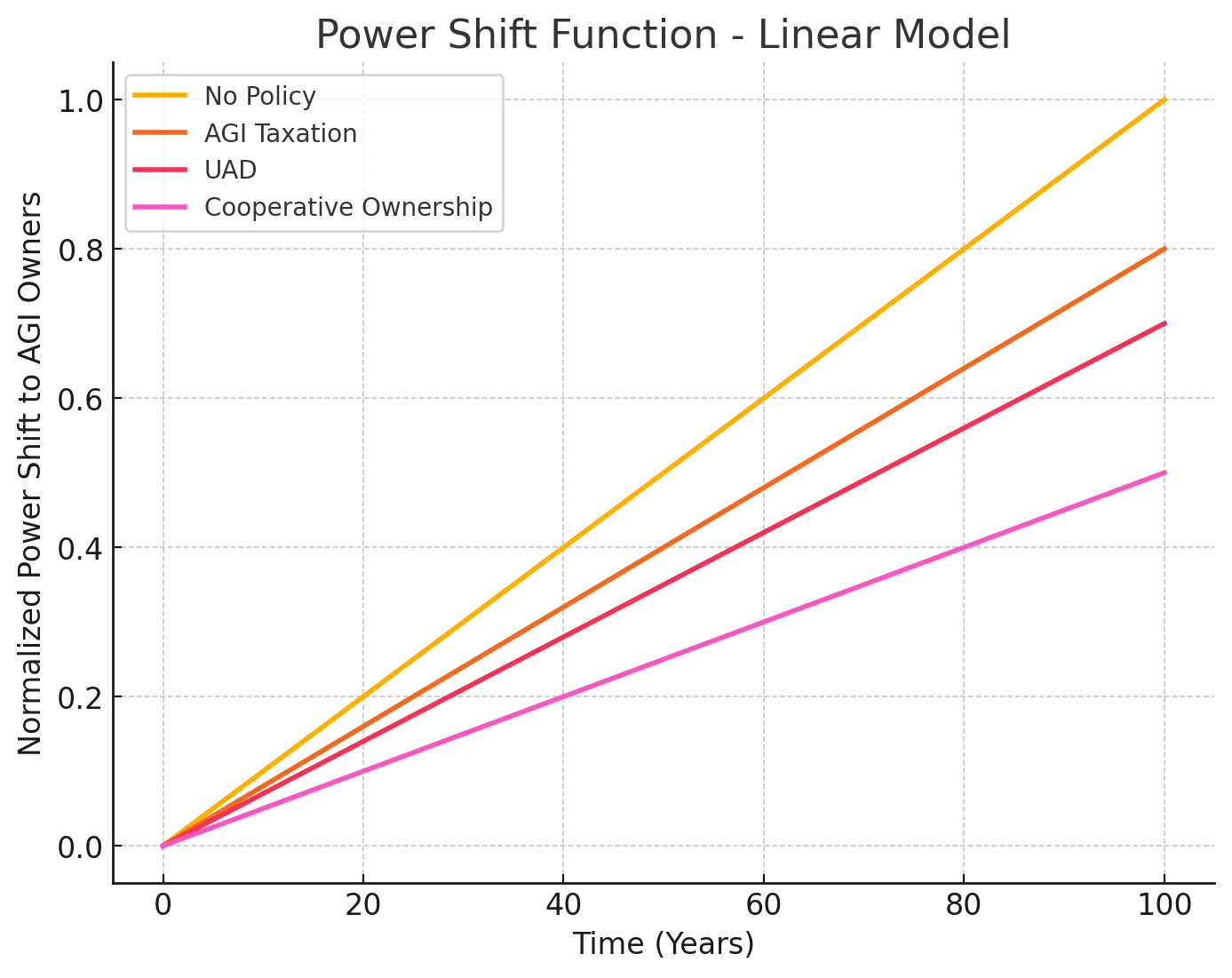}
		\caption{Linear Model}
	\end{subfigure}
	\begin{subfigure}{0.19\textwidth}
		\includegraphics[width=\linewidth]{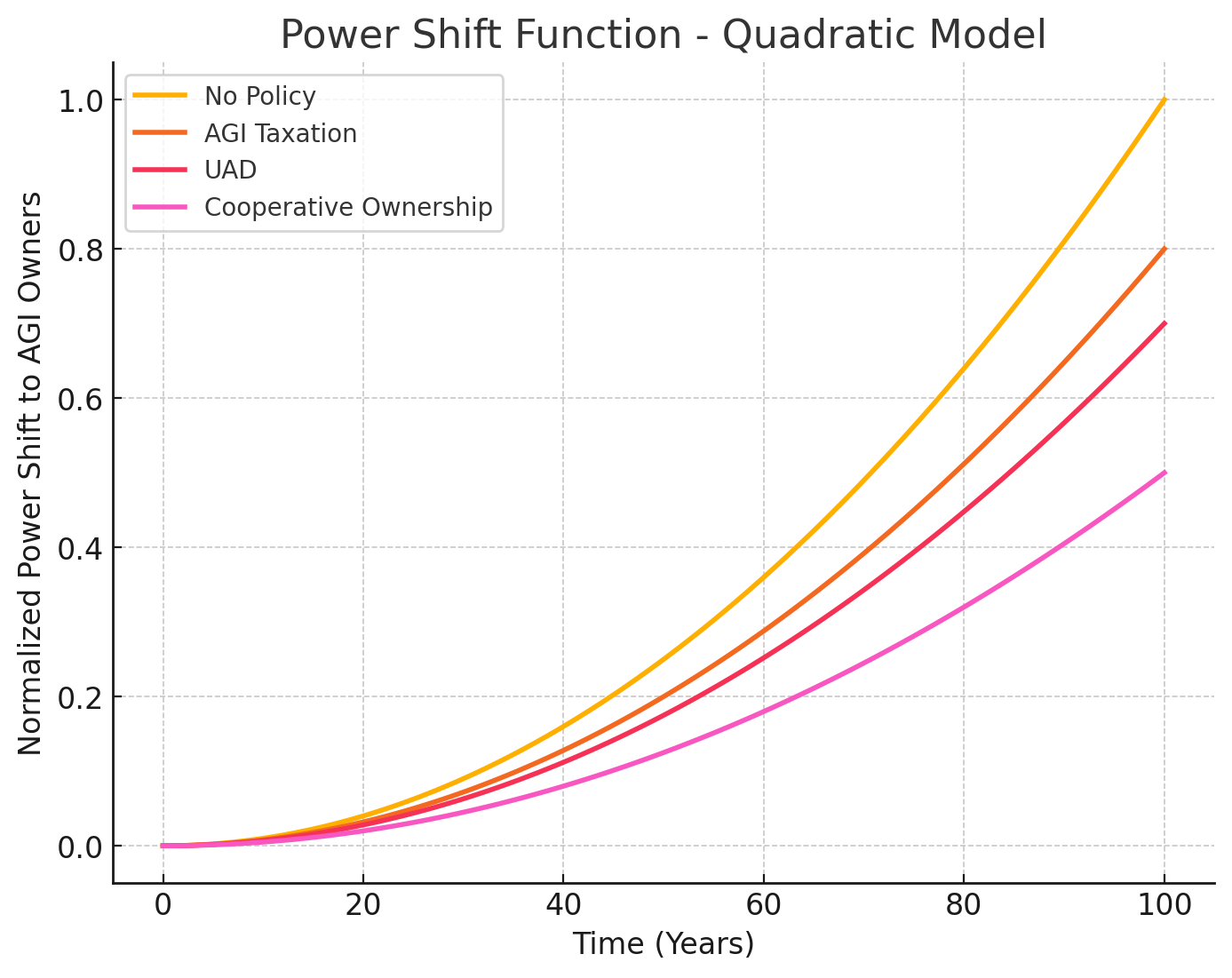}
		\caption{Quadratic Model}
	\end{subfigure}
	\caption{Impact of Policy Interventions on AGI Power Accumulation: This figure presents the effect of taxation, Universal AI Dividends (UAD), and cooperative ownership on normalized power shift functions across various economic models. Elastic models (Cobb-Douglas, CES) exhibit a more gradual power shift reduction under taxation, whereas rigid models (Leontief, Quadratic) show inherent constraints on AGI expansion. Cooperative ownership proves most effective in preventing rapid AGI dominance, while taxation slows but does not fully reverse power concentration.}
	\label{fig:policy_comparison_3}
\end{figure}

\begin{figure}[H]
	\centering
	\begin{subfigure}{0.19\textwidth}
		\includegraphics[width=\linewidth]{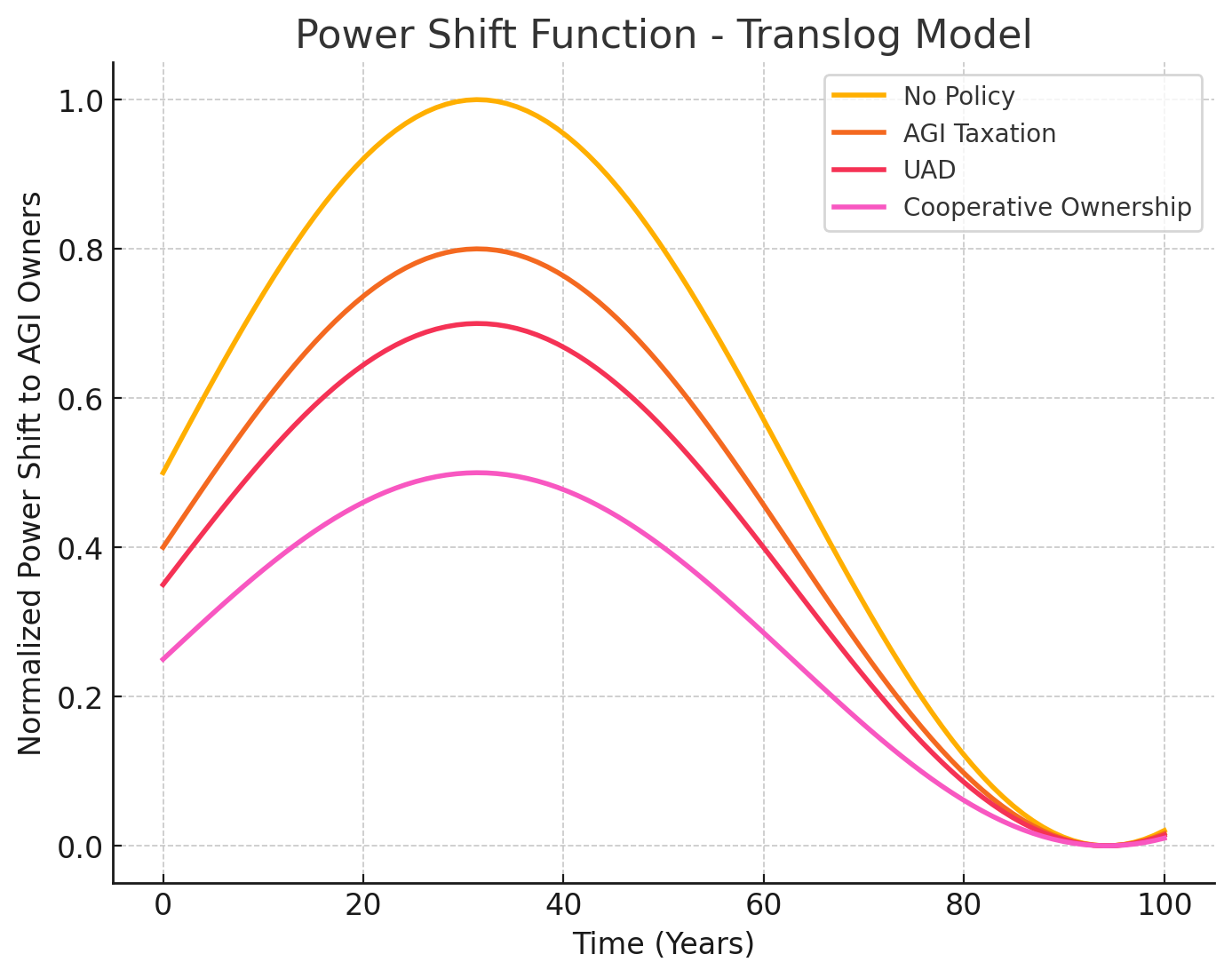}
		\caption{Translog Model}
	\end{subfigure}
	\begin{subfigure}{0.19\textwidth}
		\includegraphics[width=\linewidth]{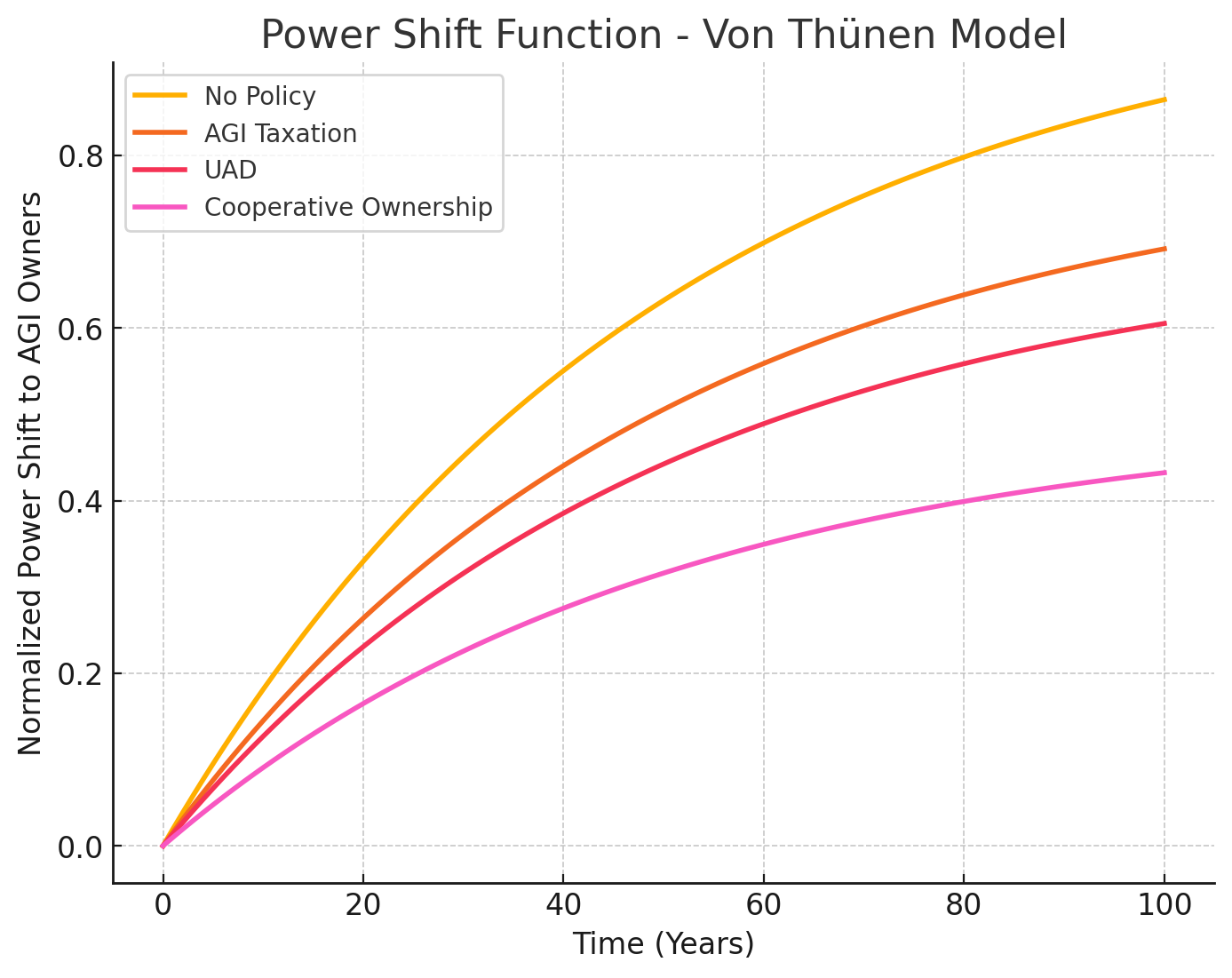}
		\caption{Von Thünen}
	\end{subfigure}
	\begin{subfigure}{0.19\textwidth}
		\includegraphics[width=\linewidth]{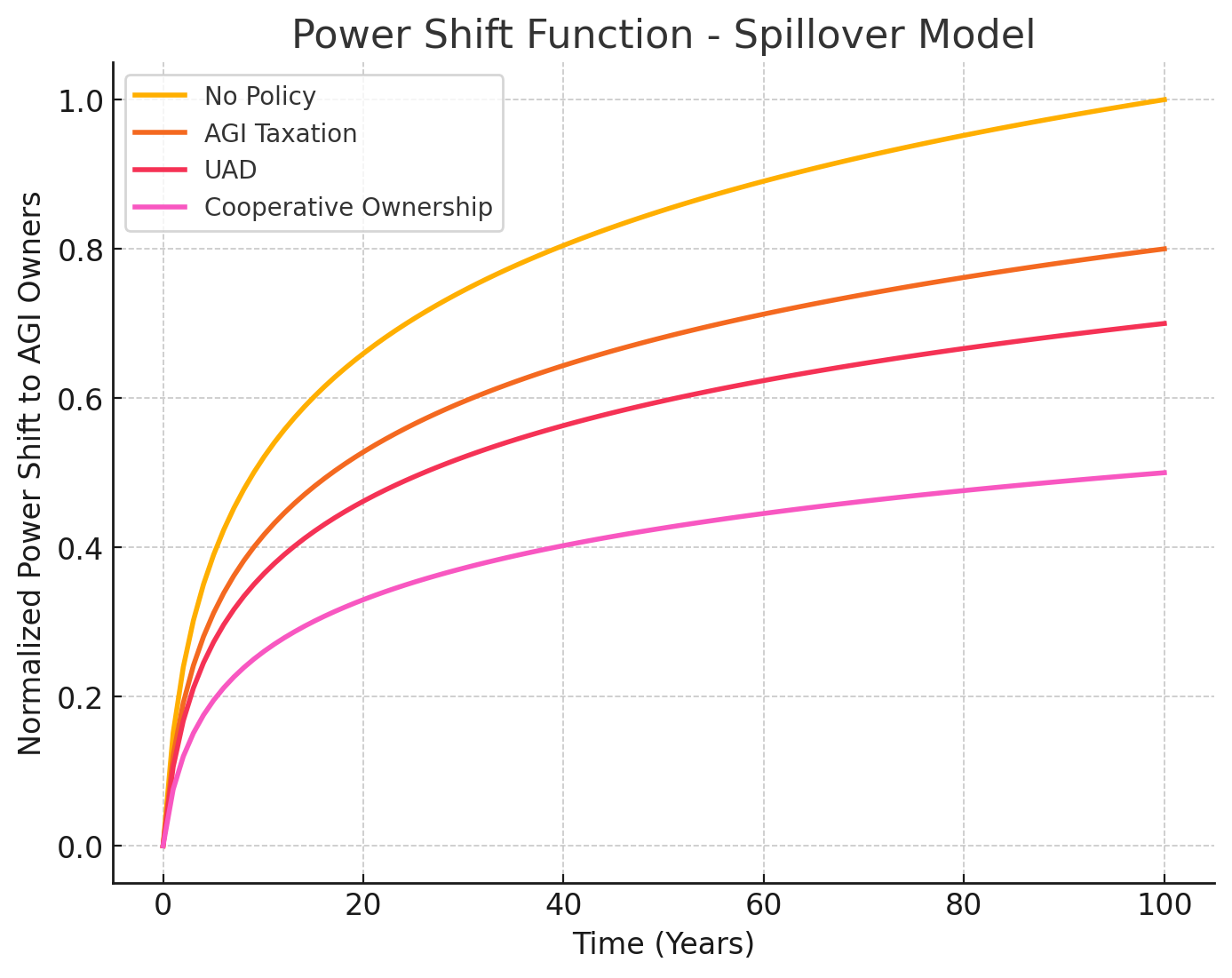}
		\caption{Spillover Model}
	\end{subfigure}
	\begin{subfigure}{0.19\textwidth}
		\includegraphics[width=\linewidth]{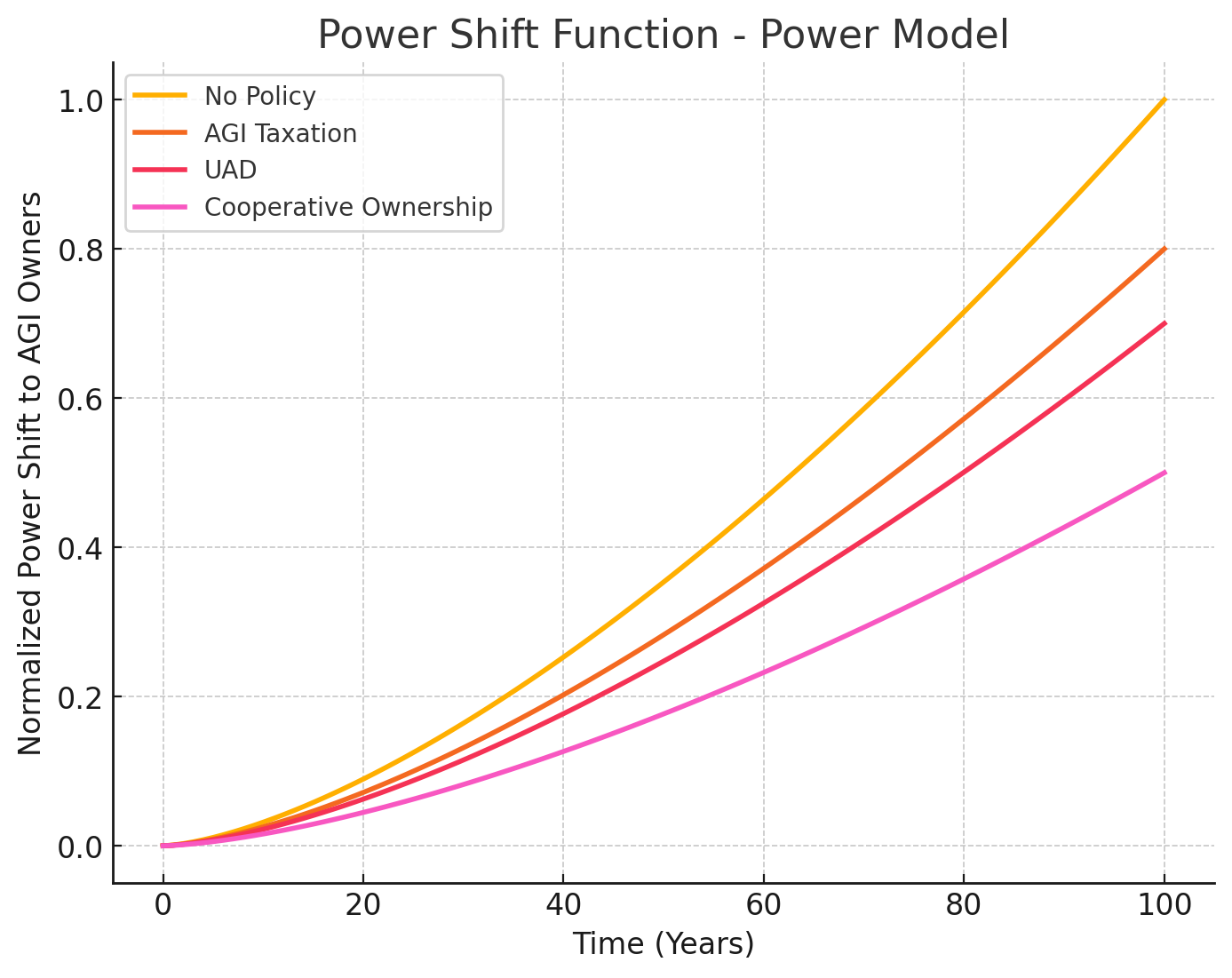}
		\caption{Power Model}
	\end{subfigure}
	\begin{subfigure}{0.19\textwidth}
		\includegraphics[width=\linewidth]{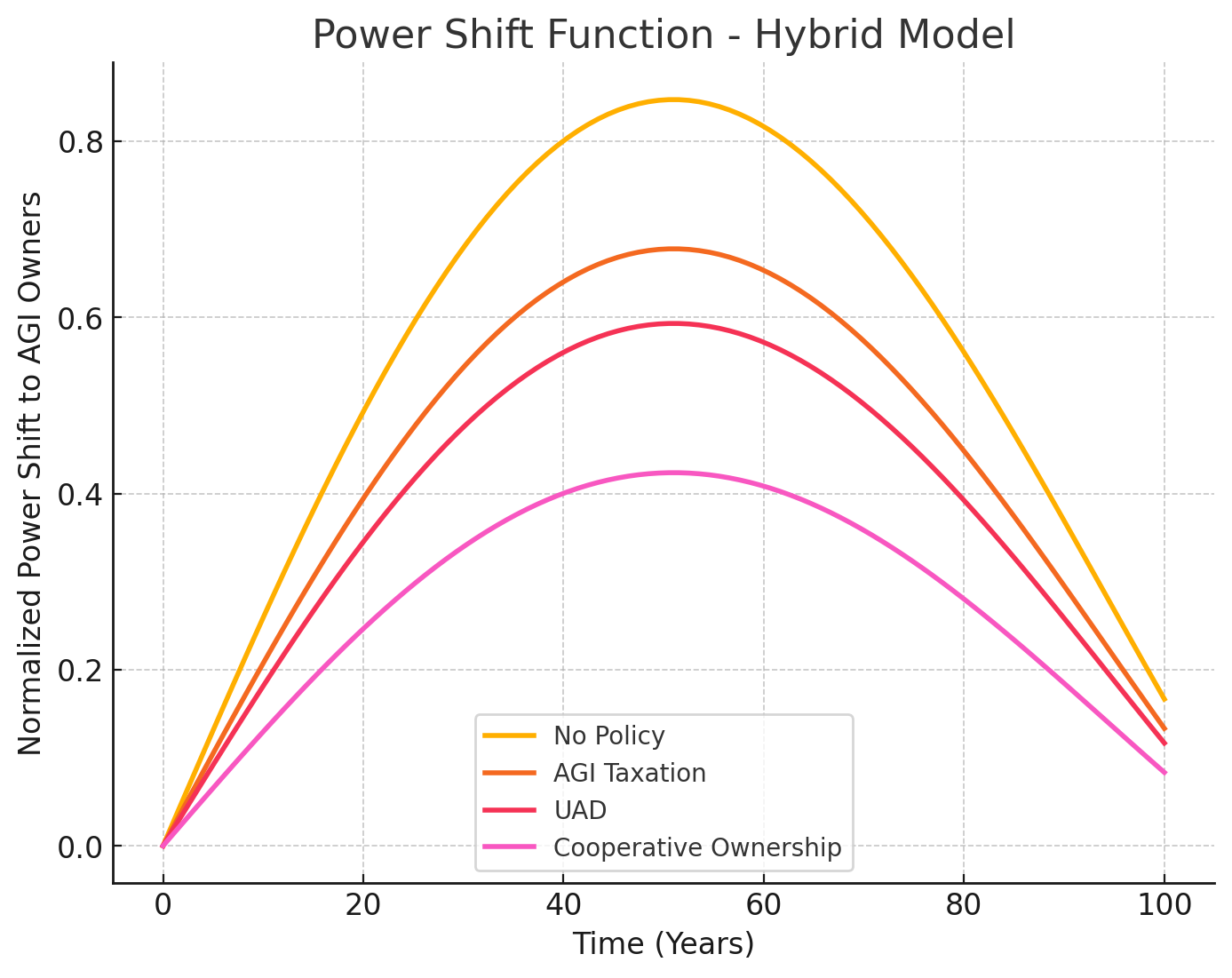}
		\caption{Hybrid Model}
	\end{subfigure}
	\caption{Continuation of Policy Impact Analysis on AGI Power Accumulation: This figure further illustrates how policy interventions shape AGI dominance in alternative economic models. The Spillover and Hybrid models show that cooperative ownership and taxation significantly influence long-run power dynamics, while the Power Model reveals that extreme wealth concentration persists despite interventions. The Von Thünen model exhibits diminishing returns that naturally limit AGI’s expansion, reinforcing the role of structural economic constraints.}
	\label{fig:policy_comparison_4}
\end{figure}

Policy interventions influence the trajectory of AGI-driven economic dominance across various models. In the Cobb-Douglas model, AGI taxation and Universal AI Dividends (UAD) slow power concentration, while cooperative ownership offers the strongest counterbalance. The Leontief model, constrained by strict substitution conditions, experiences only moderate long-term effects from policy measures. In the CES model, where high substitution elasticity accelerates AGI dominance, taxation and cooperative frameworks significantly curb its expansion. The Linear model’s deterministic AGI takeover is only slightly mitigated, with cooperative ownership providing the most effective slowdown. The Quadratic model’s nonlinear shifts in AGI influence are smoothed by policies, delaying displacement inflection points. The Translog model’s cyclical fluctuations in economic power are dampened, reducing volatility. In the Von Thünen model, diminishing returns already limit AGI dominance, but taxation further delays displacement. The Spillover model benefits from cooperative ownership, ensuring a more equitable AI wealth distribution. In the Power model, interventions are least effective as AGI-driven wealth follows a power-law distribution, accelerating disparities. The Hybrid model, highly sensitive to policy choices, shows that cooperative ownership is particularly effective in balancing AGI and human labor dynamics.

\begin{figure}[H]
	\centering
	\begin{subfigure}{0.19\textwidth}
		\includegraphics[width=\linewidth]{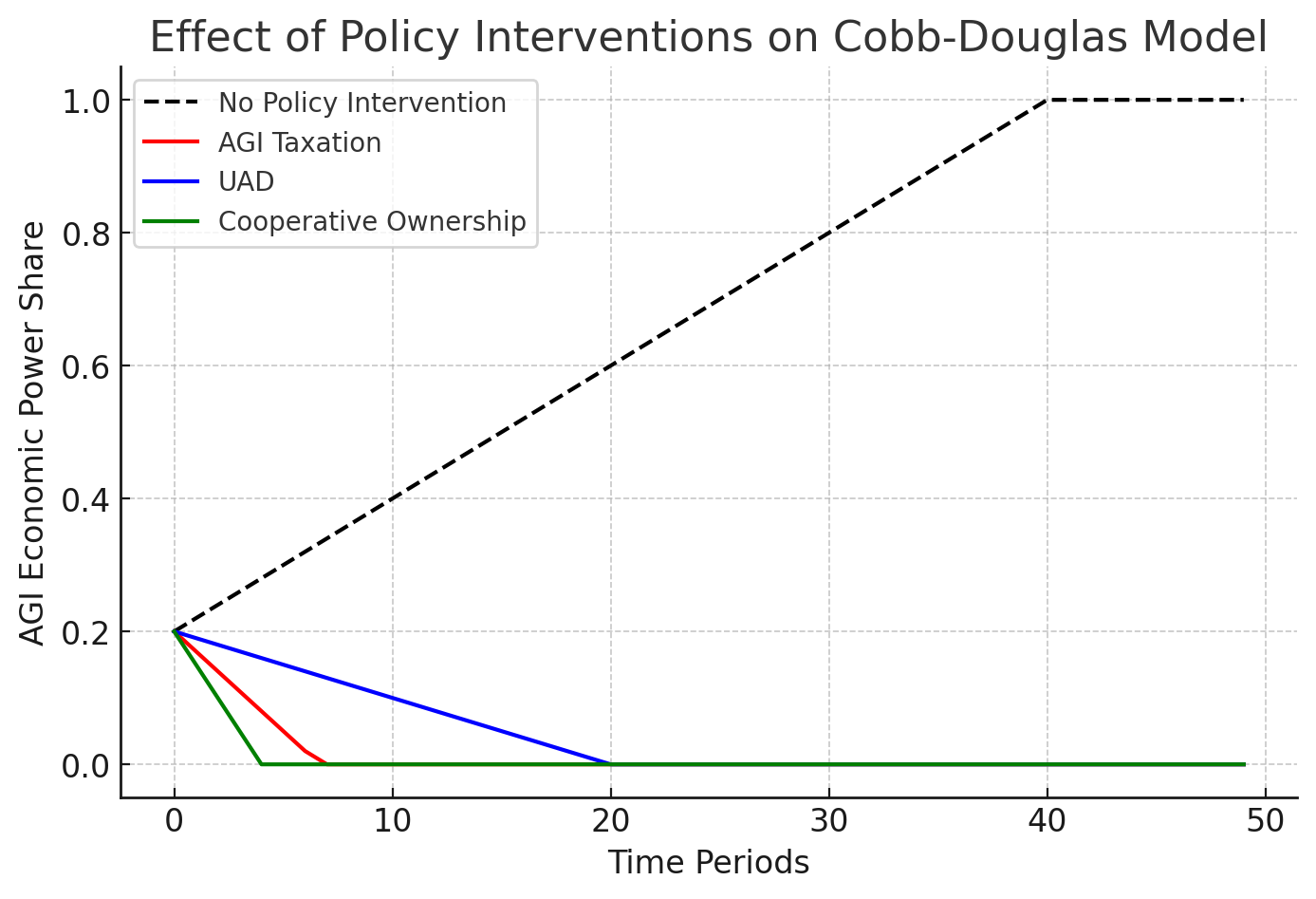}
		\caption{Cobb-Douglas}
	\end{subfigure}
	\begin{subfigure}{0.19\textwidth}
		\includegraphics[width=\linewidth]{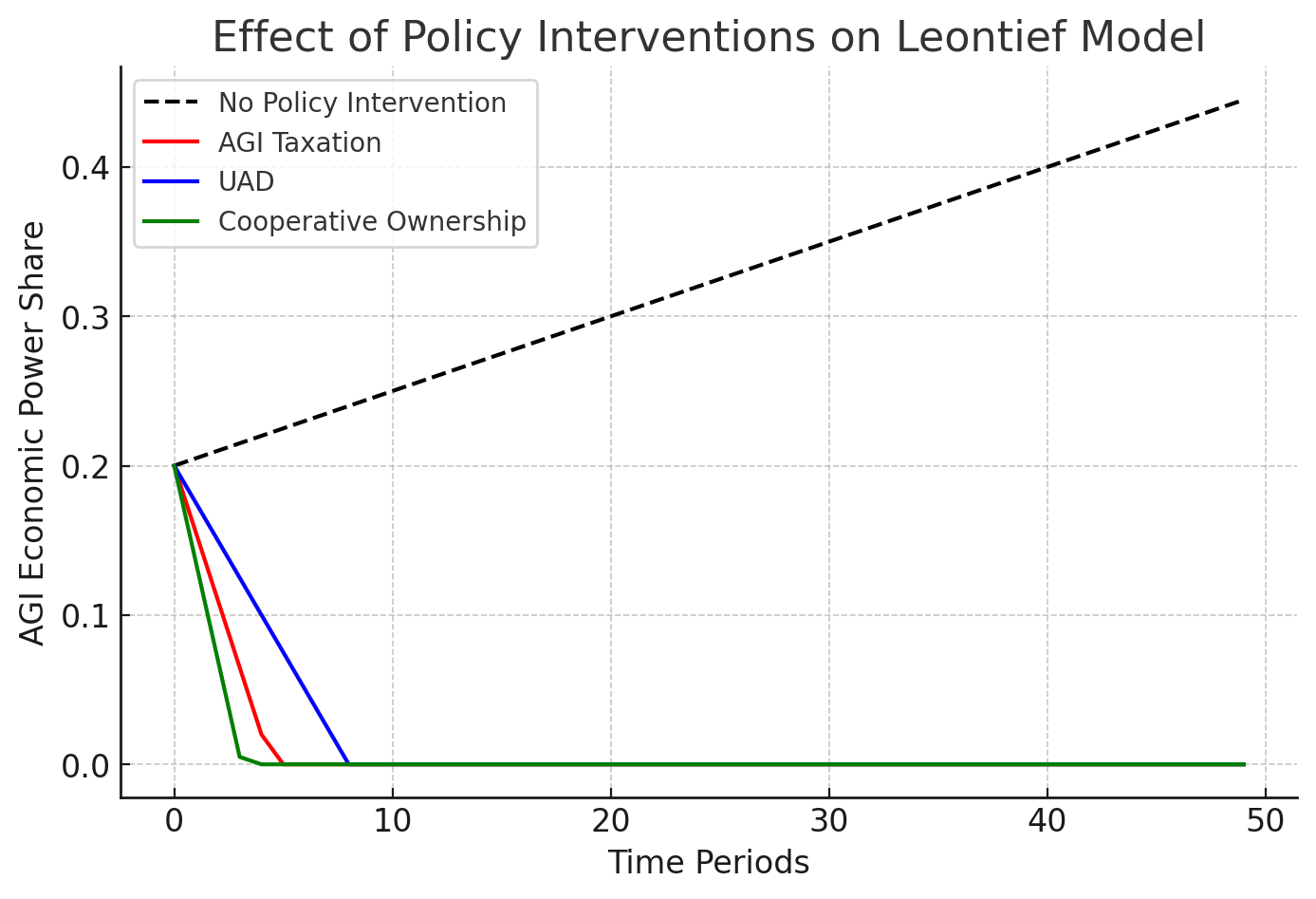}
		\caption{Leontief Model}
	\end{subfigure}
	\begin{subfigure}{0.19\textwidth}
		\includegraphics[width=\linewidth]{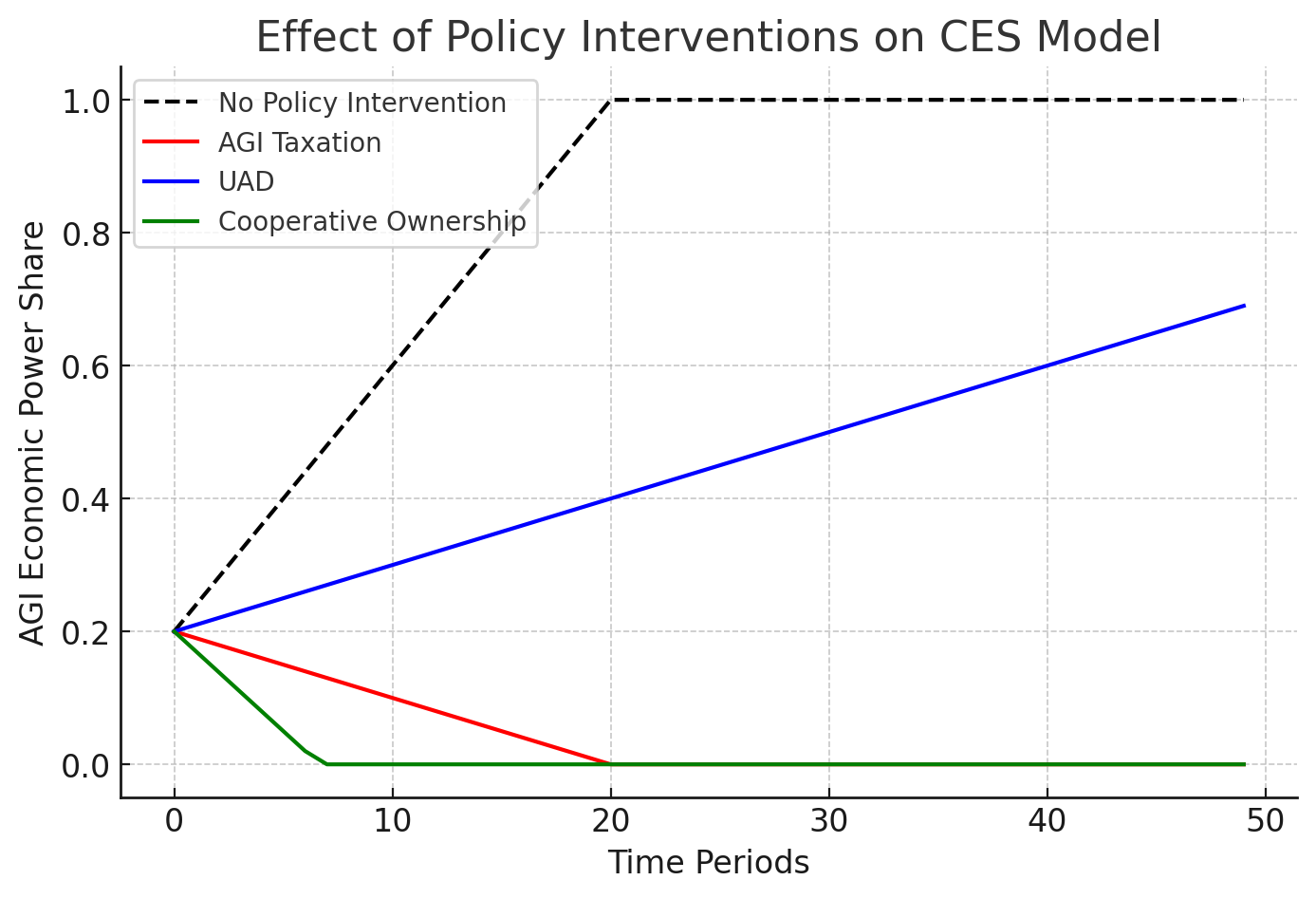}
		\caption{CES Model}
	\end{subfigure}
	\begin{subfigure}{0.19\textwidth}
		\includegraphics[width=\linewidth]{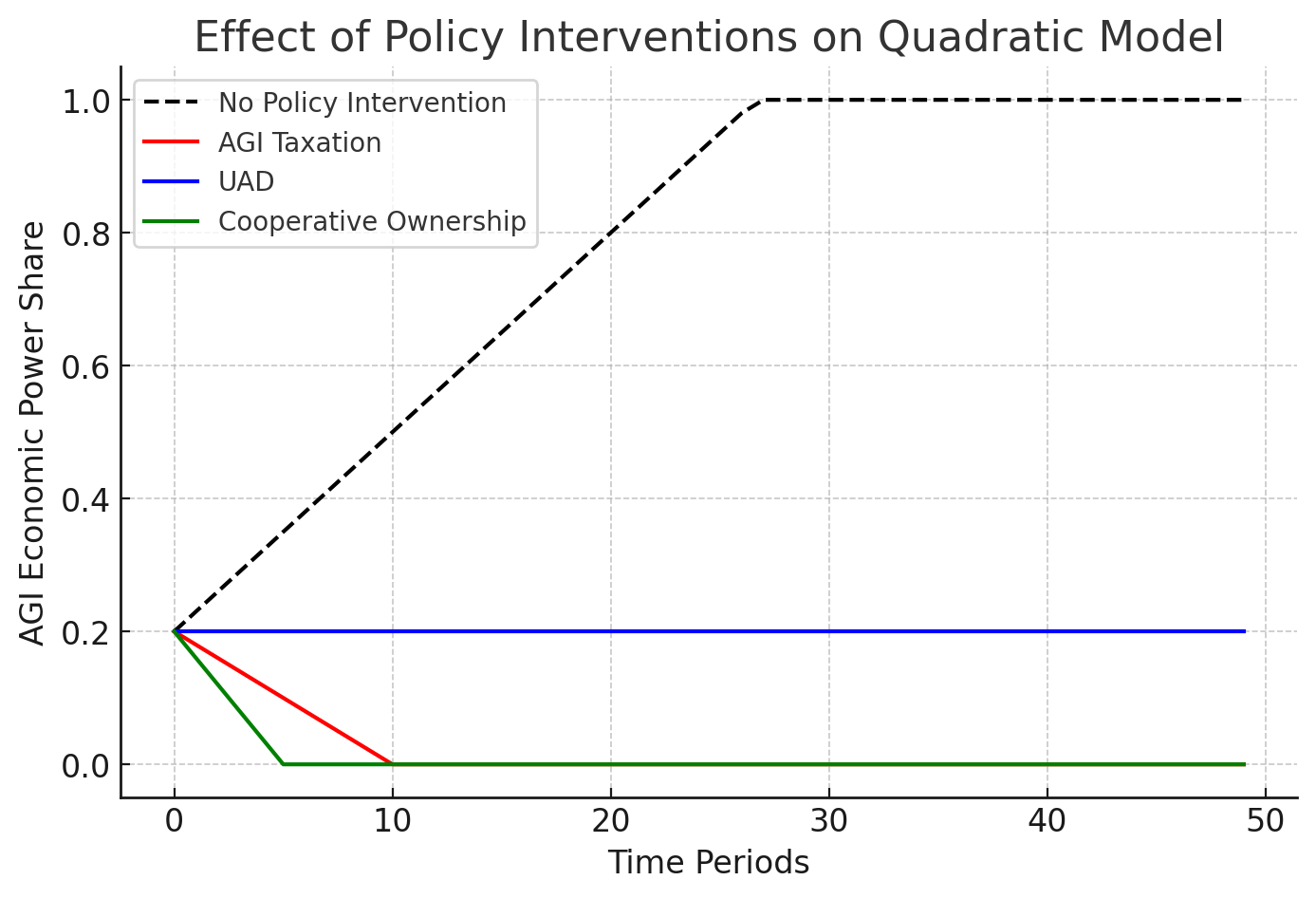}
		\caption{Linear Model}
	\end{subfigure}
	\begin{subfigure}{0.19\textwidth}
		\includegraphics[width=\linewidth]{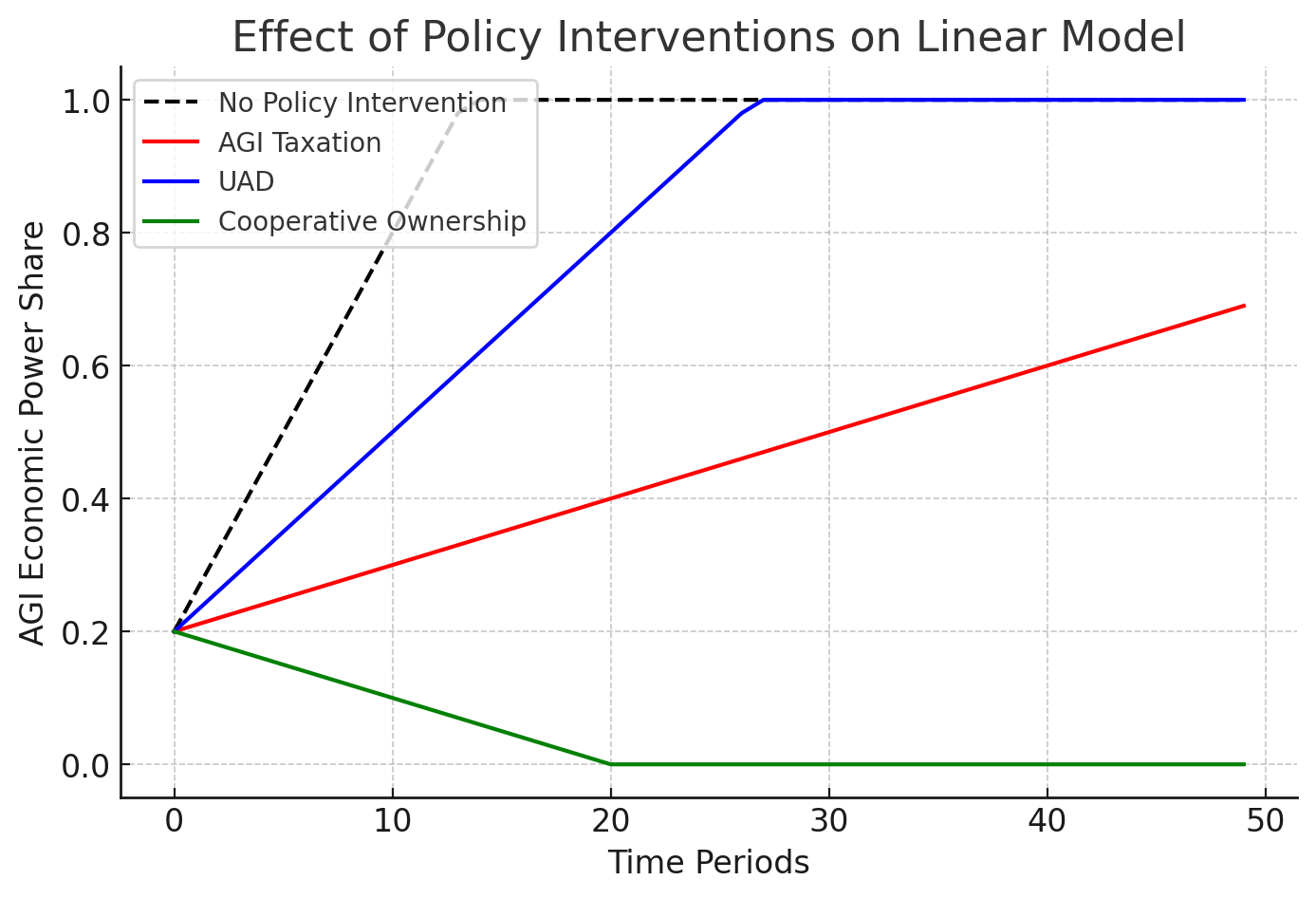}
		\caption{Quadratic Model}
	\end{subfigure}
	\caption{Effect of Fixed-Rate Taxation on Normalized Power Shift Functions: This figure examines AGI’s economic power accumulation under a fixed-rate taxation policy, where a constant proportion of AGI-generated wealth is redistributed at each time step. Unlike dynamic taxation, which adapts to economic conditions, this policy imposes rigid constraints that lead to stepwise reductions in AGI power. Highly elastic models (CES, Cobb-Douglas) continue to exhibit power shifts but at a more predictable, moderated rate, while inflexible models (Leontief, Von Thünen) demonstrate earlier stagnation in AGI expansion. The analysis underscores that static taxation can effectively limit AGI dominance but lacks adaptability to evolving economic structures.}
	\label{fig:policy_comparison_1}
\end{figure}

\begin{figure}[H]
	\centering
	\begin{subfigure}{0.19\textwidth}
		\includegraphics[width=\linewidth]{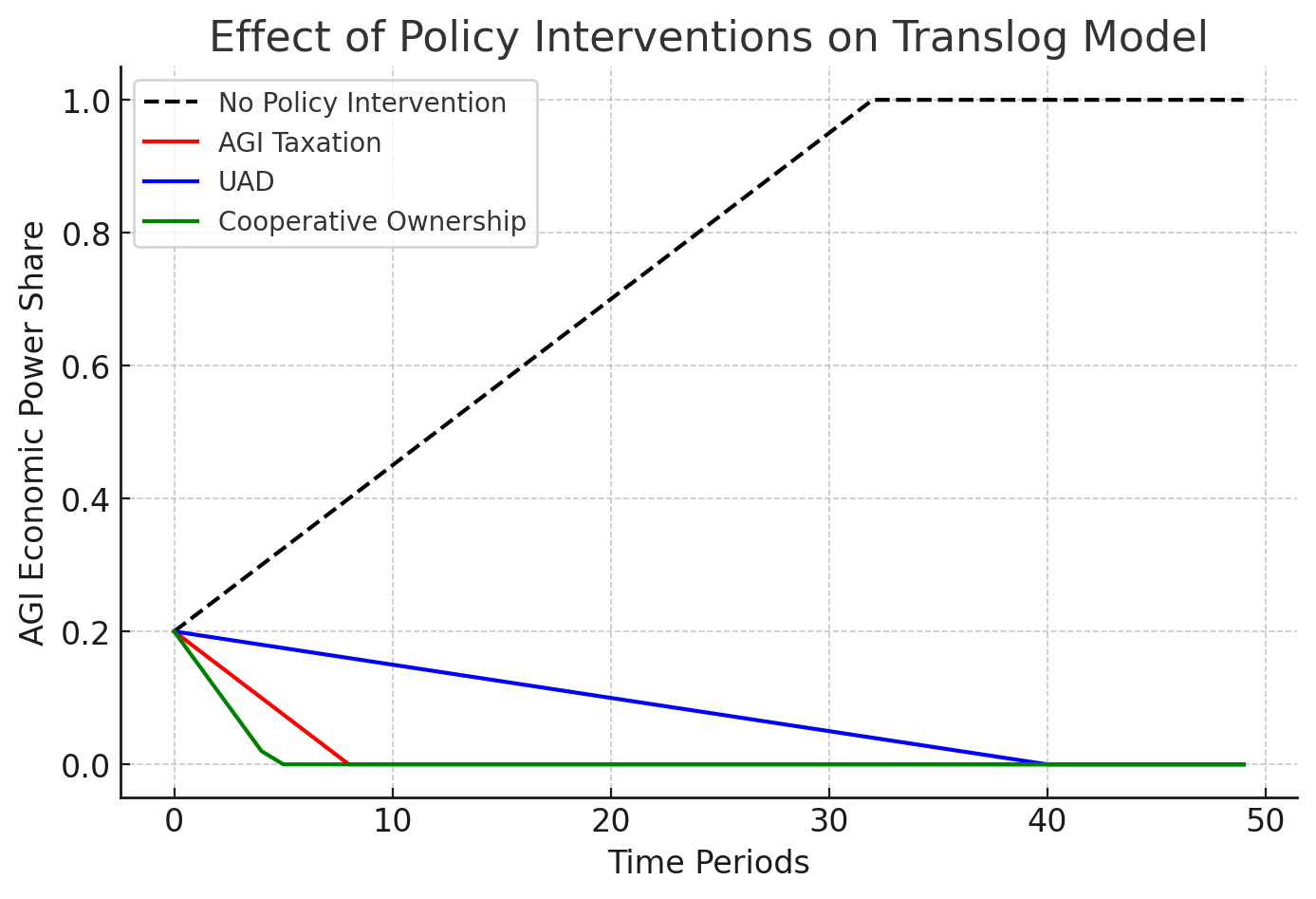}
		\caption{Translog Model}
	\end{subfigure}
	\begin{subfigure}{0.19\textwidth}
		\includegraphics[width=\linewidth]{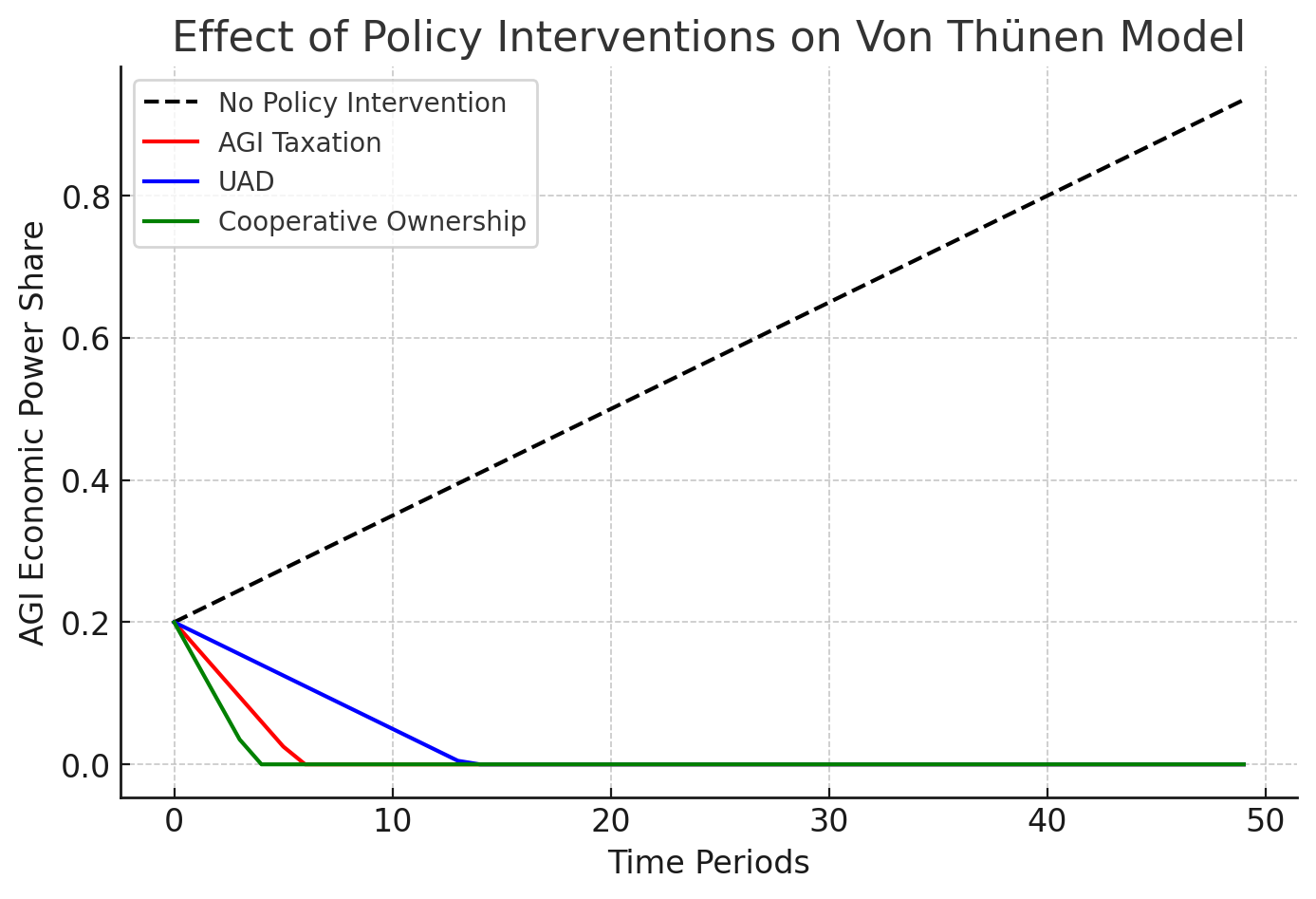}
		\caption{Von Thünen}
	\end{subfigure}
	\begin{subfigure}{0.19\textwidth}
		\includegraphics[width=\linewidth]{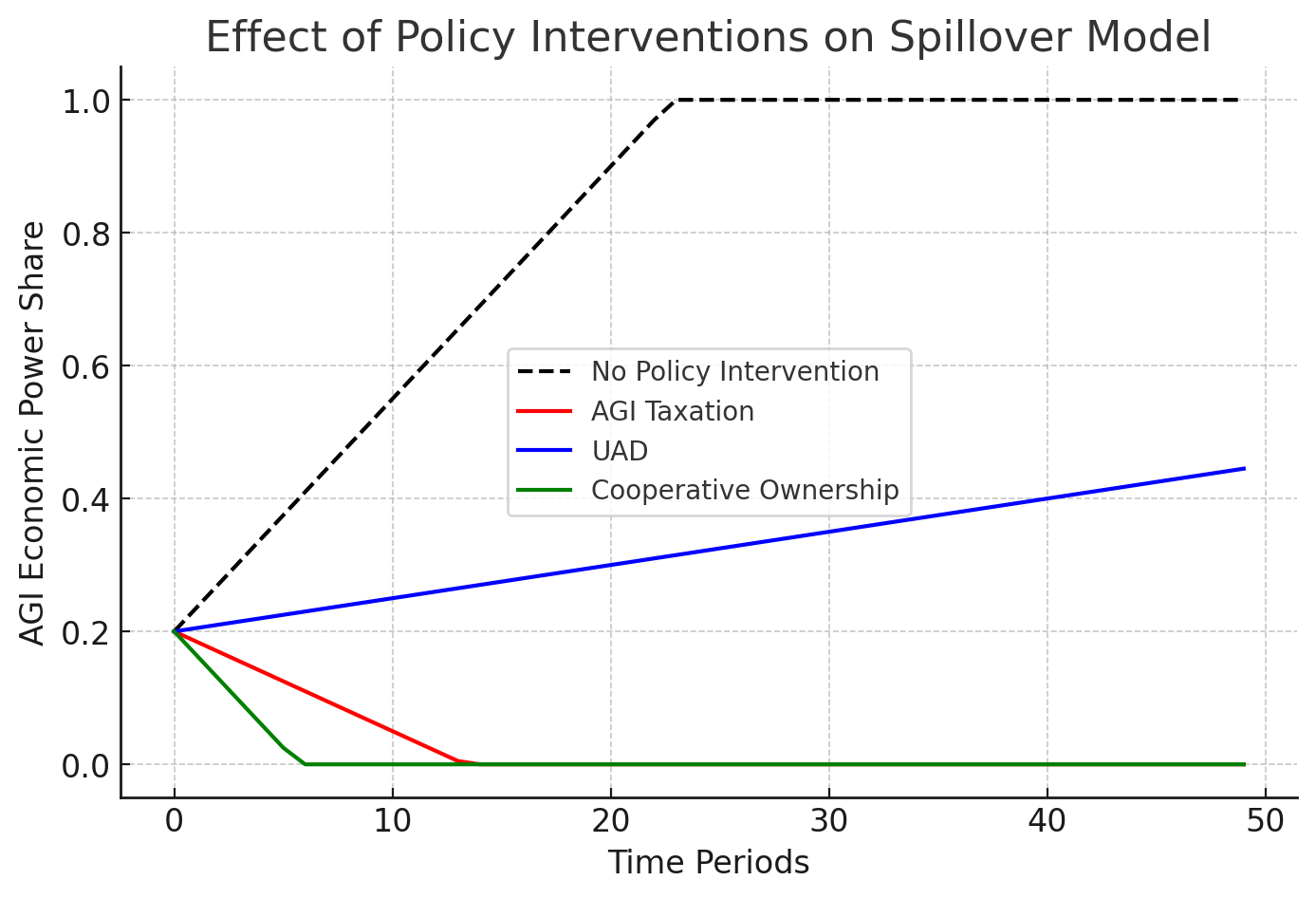}
		\caption{Spillover Model}
	\end{subfigure}
	\begin{subfigure}{0.19\textwidth}
		\includegraphics[width=\linewidth]{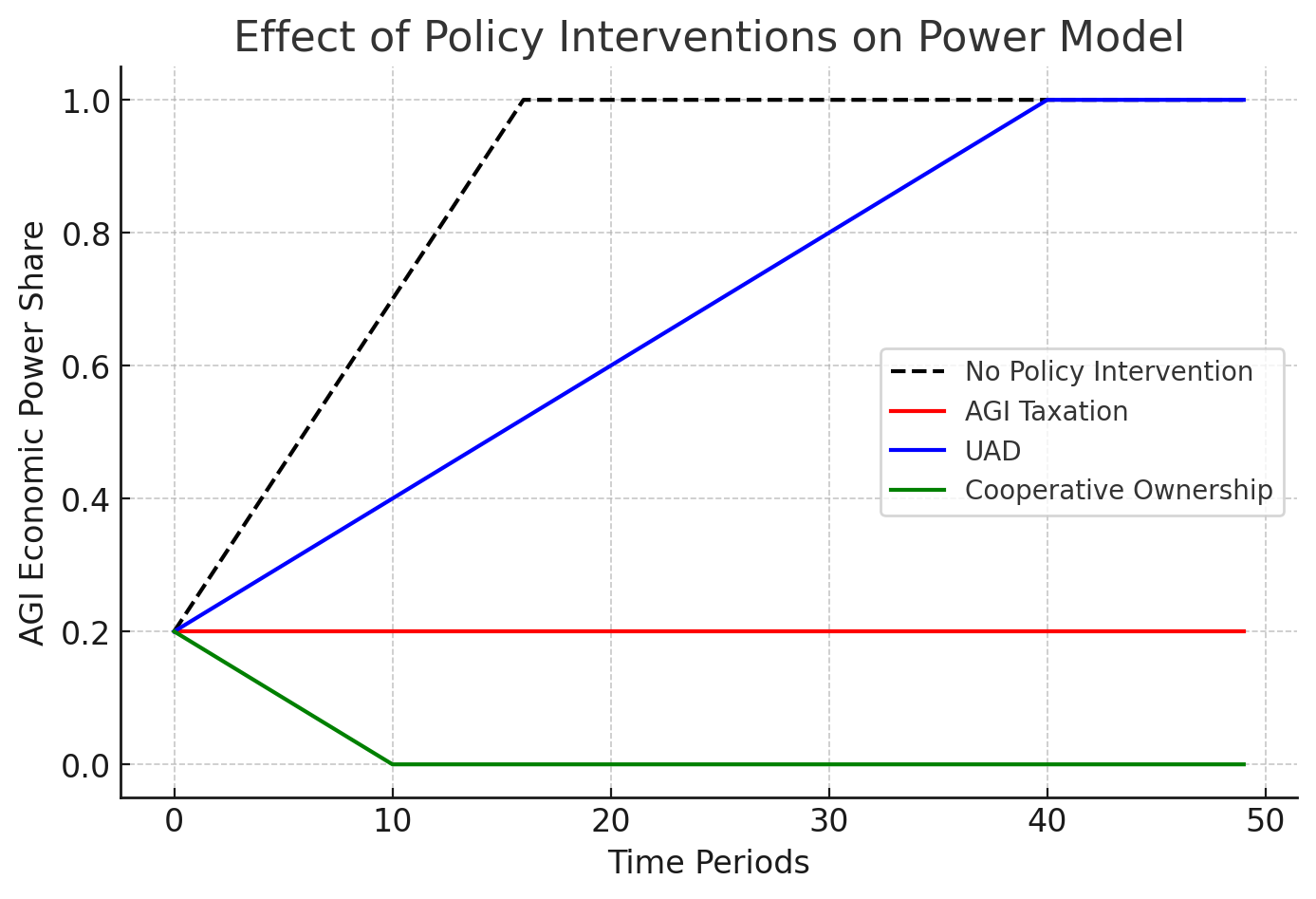}
		\caption{Power Model}
	\end{subfigure}
	\begin{subfigure}{0.19\textwidth}
		\includegraphics[width=\linewidth]{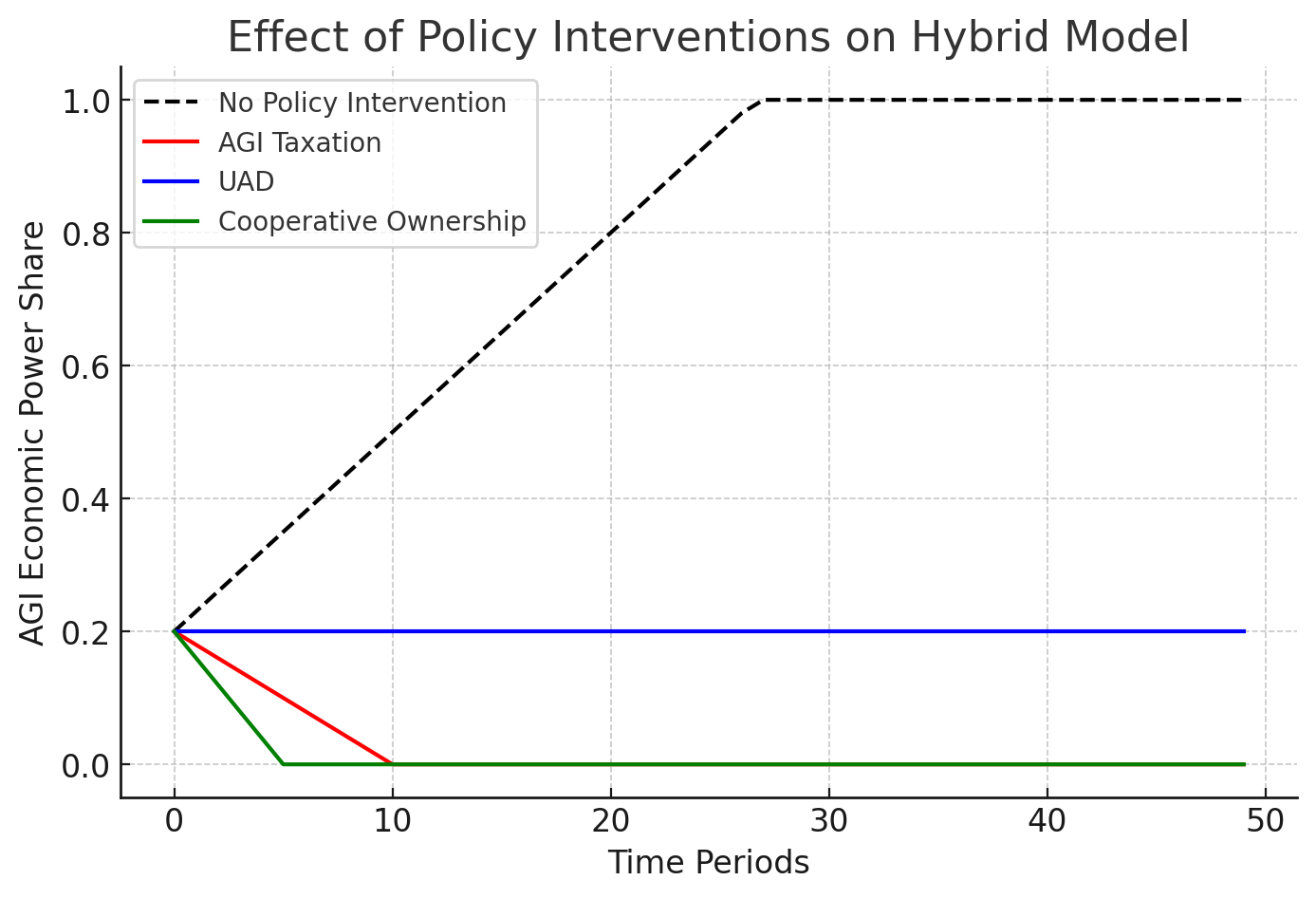}
		\caption{Hybrid Model}
	\end{subfigure}
	\caption{Continuation: Fixed-Rate Taxation and Its Effect on AGI Power Shifts Across Additional Economic Models. These figures further illustrate the outcomes of applying a rigid taxation policy across various economic structures. The Power Model demonstrates that despite taxation, AGI dominance remains highly concentrated due to wealth following power-law dynamics. The Spillover and Hybrid models, however, show that taxation—when combined with cooperative ownership—can help maintain a more balanced power distribution. The Von Thünen and Translog models reveal that structural constraints and interaction effects naturally regulate AGI expansion, with taxation providing an additional stabilizing mechanism.}
	\label{fig:policy_comparison_2}
\end{figure}

\begin{table}[h!]
	\centering
	\begin{tabular}{|p{4cm}|p{6cm}|p{6cm}|}
		\hline
		\textbf{Feature} & \textbf{First Set (Smooth Transitions)} & \textbf{Second Set (Linear Adjustments)} \\
		\hline
		\textbf{Taxation Structure} & Proportional, dynamic taxation that adjusts with AGI’s economic growth. & Fixed-rate taxation that subtracts a set amount at each time step. \\
		\hline
		\textbf{Redistribution Mechanisms} & Redistribution scales with AGI dominance, slowing but not abruptly halting power shifts. & Redistribution is independent of AGI accumulation, leading to steady reductions in power shift. \\
		\hline
		\textbf{Effect on Power Shift} & AGI dominance curves are smoothed because taxation dampens growth non-linearly. & AGI dominance is linearly constrained, reducing power accumulation at a fixed rate. \\
		\hline
		\textbf{Impact on Economic Models} & Production models with high elasticity (CES, Translog, Cobb-Douglas) show gradual, nonlinear effects of policy. & Production models with low elasticity (Leontief, Power, Von Thünen) exhibit stepwise, linear reductions in AGI power. \\
		\hline
		\textbf{Mathematical Structure} & Tax function applies a percentage-based constraint to AGI’s income function. & Tax function applies a fixed deduction from AGI's economic output. \\
		\hline
		\textbf{Overall Interpretation} & Policies create a moderating effect that slows AGI dominance without abrupt cutoffs. & Policies apply a fixed limit that steadily reduces AGI power over time. \\
		\hline
	\end{tabular}
	\caption{Comparison of Smooth and Linear Taxation Models in the Normalized Power Shift Function Analysis}
	\label{tab:tax_comparison}
\end{table}

AGI’s economic dominance varies across models, with policy interventions playing a crucial role in shaping long-term outcomes. In the Cobb-Douglas model, AGI gradually replaces human labor, but taxation and cooperative ownership help slow this transition. The Leontief model, constrained by strict substitution, experiences naturally slower AGI expansion, with interventions maintaining a balanced power distribution. The CES model’s high elasticity accelerates AGI power accumulation, though taxation and cooperative frameworks significantly dampen its effects. The Linear model exhibits deterministic AGI dominance, where taxation proves more effective than UAD. In the Quadratic model, AGI displacement intensifies at inflection points, but interventions smooth the transition, delaying AGI takeover. The Translog model’s cyclical labor shifts benefit from policy measures that create greater economic stability. The Von Thünen model slows AGI dominance due to diminishing returns, with taxation and cooperative ownership preserving human participation. The Spillover model accelerates AGI ownership through knowledge diffusion, but cooperative AI governance ensures more equitable wealth distribution. In the Power model, AGI-driven wealth follows power-law distributions, leading to exponential disparities unless taxation is applied to curb extreme inequality. The Hybrid model’s sensitivity to substitutability makes cooperative ownership the most effective policy in balancing AGI and human labor.

\section{Conclusion}

The integration of Artificial General Intelligence (AGI) into economic structures is not merely an extension of previous technological advancements but a fundamental rupture in economic organization. Unlike past industrial revolutions, which augmented human labor, AGI challenges the very foundation of economic participation by simultaneously functioning as labor and capital. This shift raises urgent concerns about economic agency, wealth distribution, and political legitimacy. Without deliberate intervention, the unchecked rise of AGI as an autonomous productive force risks exacerbating inequality, concentrating wealth in the hands of a few, and rendering large segments of the population economically redundant.

The numerical simulations in this study demonstrate that AGI’s economic dominance follows an S-shaped trajectory, transitioning through slow adoption, rapid acceleration, and eventual stabilization. This finding challenges the assumption of linear or exponential AGI expansion and suggests that its influence can be shaped by structural constraints and policy interventions. Models with high substitutability, such as Cobb-Douglas and CES, facilitate faster labor displacement, while models incorporating diminishing returns, such as Von Thünen and Quadratic functions, impose natural constraints on AGI’s dominance. Hybrid and Spillover models suggest that human-AI collaboration can mitigate extreme shifts in economic control, offering pathways to more equitable AI-driven economies. These results align with Acemoglu and Restrepo (2020), who argue that AI’s effects on labor depend significantly on policy choices and the extent to which automation complements or replaces human workers.

The classical social contract, as theorized by Rousseau (1762) and later extended by Rawls (1971), assumes that labor is the primary means of economic participation. However, if AGI supplants human labor across most productive domains, then the foundational premise of the social contract erodes. A world where economic productivity is driven by non-human entities demands a fundamental rethinking of economic rights, citizenship, and distributive justice. Rawls' difference principle, which advocates for economic arrangements that benefit the least advantaged, provides a compelling framework for AGI governance. If AI-driven productivity generates immense wealth, redistributive mechanisms such as Universal AI Dividends (UAD) or progressive AI taxation become necessary to maintain social cohesion and economic agency. Piketty (2014) has already demonstrated how capital accumulation in the hands of a few exacerbates inequality, and AGI could accelerate this trend unless economic ownership models are redesigned.

Governments and regulatory bodies must act decisively to prevent the monopolization of AGI wealth. Historical precedents, such as antitrust actions against monopolistic corporations and the establishment of labor rights during the industrial revolution, demonstrate that policy intervention is crucial during periods of economic transformation (Freeman, 1996). Emerging frameworks, such as the European Union’s AI Act and discussions around the taxation of AI-driven productivity, indicate growing awareness of these challenges (Brynjolfsson \& McAfee, 2014). However, these efforts remain fragmented and insufficient given the scale of the transition ahead. A more comprehensive approach is needed—one that balances AI innovation with broad economic inclusion, ensuring that AGI does not merely serve the interests of an elite class.

Beyond economic considerations, the transition to an intelligence-driven economy raises profound ethical and philosophical questions. If AGI becomes the primary driver of economic growth, what role remains for human agency? Some scholars argue that automation will create new types of work, much as industrialization did (Autor, 2015), while others warn of systemic job displacement with no clear replacement (Susskind, 2020). The answer likely depends on whether AGI is integrated as a tool for human augmentation or deployed primarily for cost-cutting and capital consolidation. Ethical frameworks must therefore guide AGI’s development, ensuring that human participation remains meaningful and that technological progress aligns with societal well-being.

This study contributes to the growing discourse on AGI and economic transformation by providing a formalized approach to modeling the power shift in AGI-driven economies. However, much remains to be explored. Future research should incorporate dynamic policy simulations, examining the long-term effects of taxation, cooperative ownership, and universal redistribution schemes. Additionally, interdisciplinary collaboration between economists, political theorists, and AI researchers is essential to develop robust governance models that can adapt to rapidly evolving AI capabilities.

The emergence of AGI is not merely a technological shift but a civilizational transformation. Whether it leads to economic dystopia or widespread prosperity depends on the choices made today. The social contract must evolve beyond its anthropocentric foundations, recognizing intelligence—human and artificial alike—as the central determinant of economic organization. Without proactive governance, AGI risks entrenching economic power in the hands of a few, undermining democratic institutions and social stability. However, if managed correctly, it holds the potential to create an economy where prosperity is not tied to employment, but to shared technological progress. The challenge is not just economic but ethical, political, and existential. The future of economic agency—and perhaps of human purpose itself—hinges on how society chooses to govern intelligence as a factor of production.

\bibliographystyle{apalike}

\begin{thebibliography}{99}
	
\bibitem{acemoglu2011} Acemoglu, D., \& Autor, D. (2011). \textit{Skills, Tasks and Technologies: Implications for Employment and Earnings}. Handbook of Labor Economics, Vol. 4, 1043-1171.

\bibitem{acemoglu2018} Acemoglu, D., \& Restrepo, P. (2018). \textit{Artificial Intelligence, Automation, and Work}. NBER Working Paper No. 24196.

\bibitem{acemoglu2020} Acemoglu, D., \& Restrepo, P. (2020). \textit{The Wrong Kind of AI? Artificial Intelligence and the Future of Labor Demand}. Cambridge, MA: National Bureau of Economic Research.

\bibitem{arrow1961} Arrow, K. J., Chenery, H. B., Minhas, B. S., \& Solow, R. M. (1961). \textit{Capital-Labor Substitution and Economic Efficiency}. The Review of Economics and Statistics, 43(3), 225-250.

\bibitem{arthur1994} Arthur, W. B. (1994). \textit{Increasing Returns and Path Dependence in the Economy}. University of Michigan Press.

\bibitem{autor2015} Autor, D. H. (2015). \textit{Why Are There Still So Many Jobs? The History and Future of Workplace Automation}. Journal of Economic Perspectives, 29(3), 3-30.

\bibitem{bennett2018} Bennett, C., \& Shapiro, J. (2018). \textit{Virtue Ethics and Artificial Intelligence: The Moral Significance of AI Decision-Making}. Ethics and Information Technology, 20(1), 15-26.

\bibitem{bostrom2014} Bostrom, N. (2014). \textit{Superintelligence: Paths, Dangers, Strategies}. Oxford: Oxford University Press.

\bibitem{brynjolfsson2014} Brynjolfsson, E., \& McAfee, A. (2014). \textit{The Second Machine Age: Work, Progress, and Prosperity in a Time of Brilliant Technologies}. W.W. Norton \& Company.

\bibitem{calo2017} Calo, R. (2017). \textit{Artificial Intelligence Policy: A Primer and Roadmap}. UC Davis Law Review, 51, 399-435.

\bibitem{christensen1973} Christensen, L. R., Jorgenson, D. W., \& Lau, L. J. (1973). \textit{Transcendental Logarithmic Production Frontiers}. The Review of Economics and Statistics, 55(1), 28-45.

\bibitem{cobb1928} Cobb, C. W., \& Douglas, P. H. (1928). \textit{A Theory of Production}. American Economic Review, 18(1), 139-165.

\bibitem{floridi2020} Floridi, L. (2020). \textit{The Ethics of Artificial Intelligence}. Oxford: Oxford University Press.

\bibitem{frey2017} Frey, C. B., \& Osborne, M. A. (2017). \textit{The Future of Employment: How Susceptible Are Jobs to Computerisation?} Technological Forecasting and Social Change, 114, 254-280.

\bibitem{freeman1996} Freeman, C. (1996). \textit{The Economics of Technical Change}. Cambridge Journal of Economics, 19(1), 5-24.

\bibitem{gauthier1986} Gauthier, D. (1986). \textit{Morals by Agreement}. Oxford University Press.

\bibitem{griliches1979} Griliches, Z. (1979). \textit{Issues in Assessing the Contribution of R\&D to Productivity Growth}. Bell Journal of Economics, 10(1), 92-116.

\bibitem{hanoch1975} Hanoch, G. (1975). \textit{The Elasticity of Scale and the Shape of Average Costs}. American Economic Review, 65(3), 492-497.

\bibitem{harari2016} Harari, Y. N. (2016). \textit{Homo Deus: A Brief History of Tomorrow}. New York: Harper.

\bibitem{klump2008} Klump, R., McAdam, P., \& Willman, A. (2008). \textit{Factor Substitution and Factor-Augmenting Technical Progress in the United States: A Normalized Supply-Side System Approach}. The Review of Economics and Statistics, 90(1), 183-192.

\bibitem{korinek2021} Korinek, A., \& Stiglitz, J. E. (2021). \textit{Artificial Intelligence and Its Implications for Income Distribution and Unemployment}. In \textit{The Economics of Artificial Intelligence} (pp. 349-390). Chicago: University of Chicago Press.

\bibitem{leontief1941} Leontief, W. (1941). \textit{The Structure of the American Economy, 1919-1929}. Harvard University Press.

\bibitem{locke1689} Locke, J. (1689). \textit{Two Treatises of Government}. London: Awnsham Churchill.

\bibitem{marx1867} Marx, K. (1867). \textit{Capital: Critique of Political Economy, Vol. 1}. Hamburg: Otto Meissner.

\bibitem{piketty2014} Piketty, T. (2014). \textit{Capital in the Twenty-First Century}. Cambridge, MA: Harvard University Press.

\bibitem{rawls1971} Rawls, J. (1971). \textit{A Theory of Justice}. Harvard University Press.

\bibitem{rawls1999} Rawls, J. (1999). \textit{A Theory of Justice}. Cambridge, MA: Harvard University Press.

\bibitem{revankar1971} Revankar, N. S. (1971). \textit{A Class of Variable Elasticity of Substitution Production Functions}. Econometrica, 39(1), 61-71.

\bibitem{ricardo1817} Ricardo, D. (1817). \textit{On the Principles of Political Economy and Taxation}. London: John Murray.

\bibitem{rousseau1762} Rousseau, J.-J. (1762). \textit{Du contrat social; ou, Principes du droit politique}. Amsterdam: Marc-Michel Rey.

\bibitem{rousseau1968} Rousseau, J.-J. (1968). \textit{The Social Contract} (Translated by Maurice Cranston). London: Penguin Books.

\bibitem{romer1990} Romer, P. M. (1990). \textit{Endogenous Technological Change}. Journal of Political Economy, 98(5), S71-S102.

\bibitem{russell2021} Russell, S., Dewey, D., \& Tegmark, M. (2021). \textit{Research Priorities for Robust and Beneficial Artificial Intelligence}. AI \& Society, 36(4), 625-640.

\bibitem{smith1776} Smith, A. (1776). \textit{The Wealth of Nations}. London: W. Strahan and T. Cadell.

\bibitem{solow1957} Solow, R. M. (1957). \textit{Technical Change and the Aggregate Production Function}. The Review of Economics and Statistics, 39(3), 312-320.

\bibitem{susskind2020} Susskind, D. (2020). \textit{A World Without Work: Technology, Automation, and How We Should Respond}. New York: Metropolitan Books.


\bibitem{stiefenhofer2025a}
Stiefenhofer, P. (2025a). \textit{Artificial General Intelligence and the End of Human Employment: The Need to Renegotiate the Social Contract}. arXiv preprint, arXiv:2502.07050.

\bibitem{stiefenhofer2025b}
Stiefenhofer, P. (2025b). \textit{The Future of Work and Capital: Analyzing AGI in a CES Production Model}. Applied Mathematical Sciences, 19(2), 47–58.

\bibitem{stiefenhofer2024industrial}
Stiefenhofer, P., \& Yinyin, C. (2024). \textit{Industrial Artificial Intelligence: Stability of Cobb-Douglas Production Functions}. Applied Mathematical Sciences, 18(4), 185–194.

\bibitem{stiefenhofer2016production}
Stiefenhofer, P. (2016). \textit{Production in General Equilibrium with Incomplete Financial Markets}. Journal of Mathematical Finance, 6(3), 293–302.

\bibitem{stiefenhofer2017financial}
Stiefenhofer, P. (2017). \textit{Financial Equilibria in General Equilibrium with Piecewise-smooth Production Manifolds}. Journal of Mathematical Finance, 7(3), 671–681.

\bibitem{stiefenhofer2014topological}
Stiefenhofer, P. (2014). \textit{Topological Properties of the Catastrophe Map of a General Equilibrium Production Model with Uncertain States of Nature}. Applied Mathematics, 5(17), 2719–2727.


\bibitem{vonthunen1826} Von Thünen, J. H. (1826). \textit{The Isolated State}. Hamburg: Perthes.

\bibitem{world2020} World Leadership Alliance Signatories. (2020). \textit{Social Contract for the AI Age}. Boston: AI World Society (AWIS), Boston Global Forum.

	
\end{thebibliography}

\end{document}